\documentclass[11pt,runningheads]{llncs}

\usepackage[margin=1in]{geometry}

\usepackage{hyperref}
\usepackage{booktabs} 
\usepackage[linesnumbered,ruled,vlined]{algorithm2e}

\SetAlFnt{\small}
\SetAlCapFnt{\small}
\SetAlCapNameFnt{\small}
\SetAlCapHSkip{0pt}
\IncMargin{-\parindent}

\usepackage{tikz}

\usepackage{makecell}

\usepackage{amsmath}

\usepackage{amsfonts} 

\usepackage[T1]{fontenc}
\usepackage{graphicx}
\usepackage{enumitem}

\usepackage[utf8]{inputenc} 
\usepackage{nicefrac}       
\usepackage{microtype}   
\usepackage{xcolor}    

\usepackage{multirow,array}

\graphicspath{{images/}}
\usepackage{csquotes}
\usepackage{color}

\usepackage[caption=false]{subfig}
\usepackage{tikz}
\usepackage{pgfplots}

\pgfplotsset{compat=1.18}

\usepackage[switch]{lineno}

\usepackage{thm-restate}

\usepackage{comment}

\SetAlgoNlRelativeSize{-1}

\allowdisplaybreaks

\begin{document}

\title{Centralization and Stability in Formal Constitutions}

\author{Yotam Gafni
}

\authorrunning{Y. Gafni} 

\institute{Weizmann Institute of Science \\
\email{yotam.gafni@gmail.com}}
\maketitle

\begin{abstract}

Consider a setting where a social choice function (SCF) is chosen to decide votes in a formal system, including votes to replace the voting method itself. Voters vote according to their ex-ante belief over what decisions are considered, and whether they prefer them to be decided by the incumbent SCF or the suggested replacement. The existing SCF then aggregates the voters' votes and arrives at a decision of whether it should itself be replaced. An SCF is self-maintaining if it can not be replaced in such fashion by any other SCF. Our focus is on the implications of self-maintenance for centralization, where we combine the approaches of Barbera \& Jackson (2004)  and Koray (2000). However, unlike Barbera and Jackson [2004], we do not generally restrict attention to \textit{anonymous} SCFs. Unlike Koray [2000], we do not restrict attention to \textit{neutral} SCFs. We present results that combine (1) optimistic, pessimistic and i.i.d. approaches with respect to voter beliefs, (2) different tie-breaking rules, and (3) different SCF domains. To highlight two of the results, (i) for the i.i.d. unbiased case with worst-case tie-breaking and Simple Games, we prove an Arrow-Style Theorem for stability: We show that only a dictatorship is self-maintaining.
(ii) With a pessimistic approach and best-case tie-breaking, we provide a tight characterization of self-maintaining Weighted Majority Games (WMGs), which are exactly those with minimal winning coalitions of size at most 2. We then consider three extensions, (i) forward-looking voters, (ii) bi-level rules, (iii) an application of our framework to expansion and retraction of voting rights.  
All in all we provide a basic framework and body of results for centralization and stability, applicable for institution design.

\keywords{Social Choice \and  Institution Design \and Blockchain Governance}

\end{abstract}
\newif\ifcomments   

\commentstrue

\newcommand{\yotam}[1]{{\ifcomments \color{red}{Yotam: #1} \fi }}
\newcommand{\ronen}[1]{{\ifcomments \color{blue}{Ronen: #1} \fi }}

\newcommand{\moshe}[1]{{\ifcomments \color{blue}{Moshe: #1} \fi }}

\newcommand{\agents}{\mathcal{N}} 
\newcommand{\ninput}{\mathcal{I}} 
\newcommand{\uu}{\bigcup_{U\in \mathbf{U^t}}U}
\newcommand{\upu}{\biguplus_{U\in \mathbf{U^t}}U}

\newcommand{\argmin}{\operatornamewithlimits{arg\ min}}

\newcommand{\nitest}[1]{\mathsf{NT}\left(#1\right)}

\newcommand{\tlabel}{t} 

\setcounter{tocdepth}{2} %

\section{Introduction}

A line of works stemming from the seminal papers of Barbera \& Jackson \cite{barbera2004choosing} and Koray \cite{koray2000self} considers stable voting rules. The idea is that a stable voting rule should maintain itself in a setting where voters may revise the voting rule. A reasonable assumption is that, over time, the voting rules likely to emerge and remain are the stable ones. \textit{Beliefs} also matter for stability: If there \textit{exist} beliefs for which an institution is stable, then for a certain society, this institution may hold. Even stronger, if an institution is stable for \textit{all} beliefs, we can trust that regardless of changes in beliefs, the institution should remain stable. 
Setting up an unstable voting rule may result in an unwanted centralization dynamics. This perspective of \textit{centralization} is the focus of our paper. 

More formally, consider the notion of \textit{self-maintaining} SCFs. Voters that are faced with deciding between two SCFs, consider their expected utility under each SCF. That is, using their beliefs of how future decision motions are drawn, they calculate the probability of their preference to be the outcome of the SCF, and they prefer the SCF where their preference is more likely to indeed be the outcome. This utility model follows Rae \cite{Rae1969}, and is the same utility model used in \cite{barbera2004choosing}. 
Importantly, for changing the voting rule it is not enough that (for example) a majority of the voters determine that they prefer to move to a different SCF: If the current SCF is not the majority rule, then aggregating their preferences does not necessarily result in deciding to move to the new SCF. For a change to happen, it must be that within the current power structure, the motion of changing to a different SCF indeed passes using the current SCF. 

Two remarks are in order. (i) Our baseline model considers voters that are \textit{myopic}, in the sense that they do not think ahead about the implications of changing the SCF. It is then natural to ask, if voters believe that any change in the SCF ultimately results in a dictatorship, might they avoid making any change at all? We consider such forward-looking voters in Subsection~\ref{sec:forward_looking}. (ii) As our baseline, we consider voters that want to maximize their probability of winning every decision. However, what if the voter is wrong about what should be the right decision?
A more nuanced world model incorporates some element of the ``Condorcet Jury'' model, where voters have independent signals over the true state of the world, and wish to make a correct decision. We consider this in our Appendix~\ref{sec:jury}.


\subsection{Related Work}

\subsubsection{Stable Constitutions}

A body of work, originating in papers by Barbera \& Jackson \cite{barbera2004choosing} and Koray \cite{koray2000self}, studies the concept of stability and self-choice of voting methods. The main focus of Barbera \& Jackson \cite{barbera2004choosing} is on qualified majority rules (ranging from Simple Majority to Unanimity). As they consider only \textit{anonymous} functions, centralization and inequality are not the main interest, but rather, it is operational: a tradeoff between effectiveness (easier to get motions through), and caution. They find that when the voting rule and the constitution (meaning the rule that decides how to change the voting rule) are the same, then simple majority is the unique stable rule (Theorem 1 of Barbera \& Jackson \cite{barbera2004choosing}). 
This is seemingly an encouraging result, as it shows democracy is stable. However, we find that without the anonymity restriction (Theorem~\ref{thm:unbiased_voters_char}), the same approach converges to a dictatorship. 

Unlike our baseline model, Barbera \& Jackson \cite{barbera2004choosing} allow for voting rules and constitutions to be different (e.g., to make a constitutional amendment, you would need a $\frac{2}{3}$ majority, but for a regular vote you need a simple majority). We find this distinction less alluring for two reasons: (i) Unless carefully constructed otherwise, a voting rule (rather than the higher-level constitution) allows for decisions on budgets, appointments, and emergency. With this power, it is possible to turn the higher-level voting rule into a dead letter. (ii) Our particular interest in this problem is motivated by DAOs, where bi-level rules are not common. However, in our extension of Section~\ref{sec:bilevel}, we derive some illuminating results for the bi-level case. 

Koray \cite{koray2000self} takes a different approach than Barbera \& Jackson \cite{barbera2004choosing}, and considers general SCFs (i.e., with their input being ordinal rankings over $m$ alternatives). Koray considers an axiom of self-selectivity that requires that for any preference profile, and with a freedom to decide tie-breaking, a voting rule should choose itself rather than an alternative rule, given that voters rank the voting rules according to their underlying rankings (and how the voting rules perform on these rankings). Together with other commonly used axioms (neutrality, unanimity), this axiom is equivalent to Independence of Irrelevant Alternatives (IIA). This, in turn, implies through Arrow's theorem \cite{arrow1950difficulty} that when there are at least $3$ alternatives, dictatorship is the only universally self-selective, neutral and unanimous rule. Importantly, later work \cite{binarySelfSelecting} shows that this implication extends to the binary case as well. 
Koray's \cite{koray2000self} notion of universal self-selectivity tightly relates to our ``pessimistic'' approach. 
Our most important divergence from Koray \cite{koray2000self} is that we avoid imposing the \textit{neutrality} axiom. The reason for our choice is that as emphasized by Barbera \& Jackson \cite{barbera2004choosing}, 
for constitution design, it is natural to consider the status-quo as different than the proposed motion. For example, all qualified majority rules are not neutral: They are biased towards deciding for the status-quo. Non-neutral SCFs indeed come up in our characterizations. For example, with 
a pessimistic approach with status-quo tie-breaking and \textit{neutral} and unanimous rules, we show non-neutral methods (e.g., the $2$-oligarchy, which is a non-neutral quasi-dictatorship), are also self-maintaining. 

Weighted Majority Games (WMGs) are a commonly-used class of decision rules where each voter $i$ is assigned a weight $w_i$ and a decision passes if the sum of supporting weights exceeds a threshold $T > \frac{1}{2}\sum_i w_i$. Azrieli \& Kim \cite{azrieliKim} study an application to WMGs of the framework of Barbera \& Jackson \cite{barbera2004choosing} (which, recall, only consider \textit{anonymous} voting rules: WMGs are not anonymous in general due to the different weights assigned to voters).
Incredibly, our result for pessimistic self-maintenance with status-quo tie-breaking identifies the exact same class as their characterization in Corollary~2. This is, of course, despite fundamental differences in the framework: They consider only independent distributions, and they compare WMGs with arbitrary SCFs outside the domain (including, even, randomized voting rules). Our approach is more principled in that it compares SCFs only within a certain domain (in this case, WMGs), and arbitrarily correlated beliefs. We also find that this characterization extends to the larger class of simple games, but \textit{not} to general Boolean functions (see the tables summarizing our baseline model results in Appendix~\ref{app:tables_summary}). We view this coincidence of characterizations as an indication of the robustness of the identified class, where different models and approaches that aim to characterize stability identify this same class.








\subsection{Our Approach and Contribution}

We start by defining voter \textit{belief} as a distribution over possible voting vectors, corresponding to the voter's assessment of future decisions' distribution. We then define the voter's ex-ante utility as the probability that a decision by the SCF, over a voting vector sampled from their belief, will match their preference. We say that an SCF $f$ is self-maintaining if for any other SCF $f'$, if voters vote according to their ex-ante preference between the two SCFs, and the vote is aggregated using $f$, then the final decision is to not replace $f$. Since tie-breaking over how voters vote in case $f$ and $f'$ have the same ex-ante utility can make a notable difference, we consider two tie-breaking rules: \textit{Worst-case}, that considers all possible voter votes in this case and may choose the worst one stability-wise, \textit{status-quo precedence} (SQP), that always votes to keep $f$. In the monotone domains we study, SQP tie-breaking is equivalent to best-case (stability-wise) tie-breaking. We view both tie-breaking rules as interesting: SQP tie-breaking is natural for the short-term stability of the system, as it is more natural for voters to lean towards the status-quo. Worst-case tie-breaking is natural for the long-term stability of the system, as over long enough time any tie-breaking outcome may unfold. 


For a comprehensive understanding of self-maintenance, we consider three approaches to beliefs:

\begin{itemize}

\item \textit{Optimistic.} In the optimistic approach, we consider an SCF to be self-maintaining if there \textit{exists} a common belief that supports it. 

\item \textit{Pessimistic.} In the pessimistic approach, we only consider an SCF to be self-maintaining if \textit{all} common beliefs support it.

\item \textit{I.i.d.} We assume either a biased or unbiased i.i.d. common belief, and consider an SCF to be self-maintaining if it is self-maintaining w.r.t. this belief. In an i.i.d. belief, each voter's preference w.r.t. a decision is an independent Bernoulli variable with some parameter $p$, where in the unbiased case $p = \frac{1}{2}$.

\end{itemize}

These approaches correspond to standard epistemic approaches in game theory: The optimistic approach corresponds to Rationalizability \cite{ratio1,ratio2}, meaning we can find a belief supporting a certain rule. The pessimistic approach corresponds to Robustness \cite{robustMechDesign}, meaning the rule can remain stable under any condition. I.i.d. is a middle ground where we assume independence, which is a natural case. We study three domains of functions: General Boolean functions, the more restricted Simple Games (that correspond to non-constant \textit{incentive-compatible} SCFs), and the natural and widely used in practice class of Weighted Majority Games (WMGs), where voters have weights, and a motion passes if and only if voters with over a threshold $T$ of the weights support it. 

Overall, the approaches, tie-breaking rules, and domains, create a Cartesian product of $18$ different settings. We do not \textit{fully} characterize every possible combination, but our characterization results are quite comprehensive, and are summarized in the tables of Appendix~\ref{app:tables_summary}.

Beyond our baseline model described above, we consider three extensions. The first two are different analytical lenses, meant to allow for more positive results, given our generally grim conclusions in the pessimistic and i.i.d. approaches' results. (i) \textit{Forward-looking voters.} With forward-looking voters, we consider SCFs to be self-maintaining not only if there is no viable coalition deviation to a different SCF, but also if a more complex condition holds: A coalition may believe that moving from SCF $f$ to $f'$ is beneficial, but this would later lead to moving to $f''$ that is undesirable to this coalition. If this holds for every coalition, then $f$ itself is stable through forward-looking reasoning. (ii) \textit{Bi-level Constitutions.} We consider the possibility of using a different SCF $\hat{f}$ for deciding whether to move away from an incumbent voting rule $f$, and provide a partial characterization. 

Lastly, in Appendix~\ref{sec:jury}, we consider an applicative extension to the problem of expansion and retraction of voting rights (studied by, e.g., Roberts \cite{roberts1999} and Acemoglu \& Robinson \cite{acemoglu2001theory}). In particular, we modify our utility function to accommodate 
``Condorcet Jury Theorem'' effects of information aggregation. We limit voting rules to a nested sequence of oligarchies, ranging from a dictatorship (oligarchy of $1$) to full democracy (oligarchy of $n$). The important parameters of this model are how much of the utility comes from taking advantage of existing resources, vs. making correct decisions ($\lambda$), and how good are voters at knowing the correct decision ($p$). We demonstrate that the stable oligarchy size is decreasing in both parameters, and that there may be multiple stable oligarchies with significant size differences. 


\section{Model}
\label{sec:model}
We consider a setting with $n\geq 3$ voters. \footnote{We focus on $n\geq 3$ to have complete characterization results (for example, the $3$-oligarchy, which appears in some of our characterizations, is undefined for $n \leq 2$). With appropriate adjustments, our results apply to the degenerate cases of $n \in \{1,2\}$ as well.}

\begin{definition}
A \textit{social choice function} (SCF) $f:\{0,1\}^n \rightarrow \{0,1\}$ is a mapping from $n$ binary preferences into a binary decision. 
\end{definition}


We now introduce a ``self-referential'' method of reasoning about SCFs to consider whether an SCF prevents itself from being changed. To allow this self-reference, it is necessary to define voter utilities w.r.t. different SCFs, and a pre-requisite for that is to define their beliefs. Our starting point is thus defining beliefs:

\begin{definition}[Voter Belief.]
 A voter belief is a probability distribution $F$ over the discrete space $\Omega = \{0,1\}^n$. I.e., for any $v \in \Omega$, 
$$ Pr_{\hat{v}\sim F}[\hat{v} = v] \geq 0,$$
and 
$$\sum_{v \in \Omega} Pr_{\hat{v}\sim F}[\hat{v} = v]  = 1.$$

\end{definition}

\begin{remark}
We choose to talk about ``beliefs'' rather than ``individual preferences'', but it is possible to assign the belief formulation an interpretation where it encodes a voter's normalized ex-ante preferences over different possible decisions. For example, if the voter knows of an upcoming important decision, where it knows all the voters' YES/NO preferences, then its belief can assign the full $1$ weight over this voting vector, regardless of whether it knows of other upcoming votes. In this sense, it makes sense to talk about individual and incompatible beliefs.
However, our results consider common beliefs, where all voters have the same distribution. This is both a natural assumption, but also aligns well with our results: In the optimistic case, common belief is a stricter requirement (and still, we show that all ``reasonable'' SCFs are self-maintaining). For the pessimistic case, common belief is an easier requirement, but still we show that almost no SCFs are self-maintaining. 
\end{remark}

\begin{definition}
Let $F$ be the voter's belief. 

Voter $i$'s \textit{ex-ante utility} of an SCF $f$ takes the following form:

$$u_i(f, F) = E_{v \sim F}[ 1[f(v) = v_i]] = Pr_{v \sim F} [f(v) = v_i]. $$

Voter $i$'s preferred choice among two SCFs $f_0, f_1$ is:

\begin{equation}
\label{eq:voter_choice}
c_i(f_0, f_1, F) = \arg \max_{j \in \{0, 1\}} u_i(f_{j}, F) \end{equation}

If $F$ is clear from the context, we simply write $u_i(f)$ and $c_i(f, f')$. 

\end{definition}


The above rule for voter's choice (Eq.~\ref{eq:voter_choice}) does not clarify how to deal with the case when $u_i(f_0) = u_i(f_1)$. We consider two possible approaches to it.

\begin{itemize}
\item \textit{Worst-case Tie-Breaking.} With this approach, a voter may vote either For or Against a motion when they are indifferent to the winner.   

\item \textit{Status-Quo Precedence (SQP).} With this approach, a voter votes Against a motion when the voter is indifferent to the winner. 

\end{itemize} 


Notice that having $c_i(f_0, f_1)$ for every voter naturally defines a vector function $\mathbf{c}(f_0, f_1) = (c_1(f_0,f_1), \ldots, c_n(f_0,f_1))$. To be compatible with prior notation, we treat $c$ as a function from $\{0,1\}^n \rightarrow \{0,1\}^n$, by encoding $f_0$ as $0$ and $f_1$ as $1$. 

\textit{Worst-case tie-breaking} is more \textit{restrictive} w.r.t. which functions are considered self-maintaining. That is because voters with indifference in their preferences have the freedom to vote in different ways, and thus \textit{expand} the space of choice outcomes, which in turn increases the possibility of coinciding with such preference vectors that result in changing the voting rule. As we will see, indeed there are settings with voting rules that are self-maintaining under \textit{status-quo precedence} but not with \textit{worst-case} tie-breaking. We also note that for \textit{monotone} domains (defined immediately below), SQP tie-breaking is best-case tie-breaking (w.r.t. stability), as it leans most to maintaining the incumbent. 

\begin{definition}[Monotone SCFs and Domains]

    For a set of voters $S$ let $v(S)$ be the voting preference where $v_i = 1[i \in S]$. We say an SCF is \textit{monotone} if for any two $v^1, v^2 \in \Omega \stackrel{def}{=} \{0,1\}^n$ so that $v^1 = v(S_1), v^2 = v(S_2)$, and $S_1 \subseteq S_2$, if $f(v^1) = 1$ then $f(v^2) = 1$. I.e., a decision may not be rejected because \textit{more} voters vote for it. 

    We say a domain is monotone if all SCFs in the domain are monotone. 
\end{definition}

We consider three domains of social choice functions:

\begin{definition}
\leavevmode 

\begin{itemize}

    \item \textit{General Boolean SCFs ($D_G$).} These include all SCFs $f:\{0,1\}^n \rightarrow \{0,1\}$, without further qualification.

    \item \textit{Simple Voting Games ($D_S$).} These include all \textit{monotone, unanimity-preserving} general Boolean SCFs. I.e., for any voter $i$, and preferences $\mathbf{v}_{-i}$, monotonicity: $f(1, \mathbf{v}_{-i}) \geq f(0, \mathbf{v}_{-i})$, and unanimity-preserving: $f(\mathbf{0}) = 0, f(\mathbf{1}) = 1$ hold. 

    \item \textit{Weighted Majority Games ($D_W$).} These include all general Boolean SCFs that can be represented using linear weights and a majority threshold. Formally, there are such $w_1, \ldots, w_n, T \in \mathbb{N}$, so that $\frac{1}{2} \sum_{1\leq i\leq n} w_i \leq T < \sum_{1\leq i\leq n} w_i$, and 
    for any $v\in \{0,1\}^n$, $f(v) = 1\left[ \sum_{1\leq i\leq n} 1[v_i = 1] \cdot w_i > T \right]$. 
\end{itemize}
    
\end{definition}

The cooperative games literature has studied and characterized these different domains extensively. In particular, it is known that $D_W \subseteq D_S \subseteq D_G$, and that a non-constant Boolean SCF is incentive-compatible if and only if it belongs to $D_S$. A useful notion in monotone games is that of a \textit{minimal winning coalition}. 
Recall that $v(S)$ is the voting preference where $v_i = 1[i \in S]$. A minimal winning coalition of a monotone SCF $f$ is a set $S$ of voters so that $f(v(S)) = 1$, but for any strict subset $S' \subset S$, $f(v(S')) = 0$ (and by monotonicity, every strict superset $S \subset S'$, $f(v(S')) = 1$). We similarly define maximal losing coalitions of a monotone SCF $f$ as a set $S$ so that $f(v(S)) = 0$ but any strict superset $S'$ has $f(v(S')) = 1$. 

\begin{definition}
\label{def:self_maintaining}
We say an SCF $f \in D$ is self-maintaining for belief $F$ and domain $D$ if for any other SCF $f' \in D$, and any choice $\mathbf{c}$ that is compatible with voter utilities towards $f, f', F$ and the tie-breaking rule we consider,

$$f(\mathbf{c}(f, f', F)) = 0,
$$

i.e., any motion to replace $f$ with another function $f'$, where voters vote according to their ex-ante preference, and the vote is aggregated using $f$ fails.

\end{definition}

Our self-maintenance condition is asymmetric in nature, in that it fixes the label for maintaining $f$ at $0$ and to replace it at $1$. 
Usually with social choice alternatives, neutrality w.r.t. the labels makes sense, since the labeling of candidates is arbitrary, and indeed notable theorems such as May's theorem use this condition. However, since our motivating setting is DAOs, or alternatively, government forms in societies, there is a natural distinction between the status quo (the current inner state of the smart contract, the current set of laws), and the suggested change. Thus, non-neutral SCFs such as \textit{Unanimity} make sense in this setting. To allow that, we do not impose neutrality.

\begin{figure}[htbp]
    \centering
    \includegraphics[width=\textwidth]{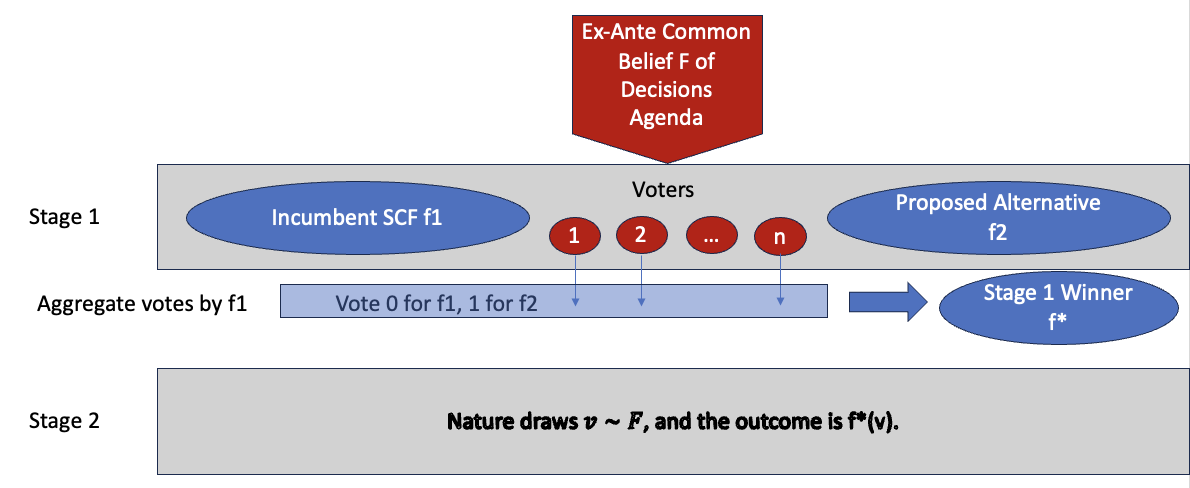}
    \caption{A game description of the self-maintaining property for belief $F$. An SCF $f_0$ is self-maintaining for belief $F$ if it wins the vote of stage $1$ as the incumbent against any SCF $f_1$.}
\label{fig:self_maintaining_game}
\end{figure}

\begin{remark}[Proof technique for handling the different domains]
Between our three domains of interest, results do not transfer without special care. For example, assume we characterize the SCFs in $D_G$ that are self-maintaining w.r.t. belief $F$, and we wish to use this for a characterization of the SCFs in $D_S$ that are self-maintaining w.r.t. $F$. We face two issues: (i) Some of the SCFs of the original characterization may not be in $D_S$. For this, we can just prune out the SCFs of interest. 
(ii) There may be SCFs in $D_S$ that are self-maintaining w.r.t. $F$ and domain $D_S$, but not w.r.t. the larger domain $D_G$. 
To avoid (ii), we use \textit{domain-preserving} constructions when ruling out an SCF. A domain-preserving construction means that if for a given $f$, we find such $f'$ that shows $f$ is not self-maintaining w.r.t. $F$, we ensure that $f'$ preserves the domain of $f$. I.e., if $f$ is in $D_S$, $f'$ is also in $D_S$, and if $f$ is in $D_G$, then $f'$ is also in $D_G$. This allows our proofs to serve all three domains at once.  
Concretely, this means that we show, for every construction of such $f'$ in our proofs, (i) It preserves monotonicity and unanimity-preserving: If the original $f$ is monotone and unanimity-preserving, then so is $f'$. This ensures it is domain-preserving for $D_S$. (ii) It preserves $D_W$: If the original $f$ is a weighted majority game, then $f'$ can also be represented using some valid weights and threshold. 
\end{remark}

An operation that is useful in this regard is looking at \textit{minimal} winning coalitions. For a monotone SCF $f$, if $f(v) = 1$, and we let $S(v) = \{i \text{ | } v_i = 1\}_{1\leq i\leq n}$, then we can find such $v'$ so that $S(v') \subseteq S(v)$ and $f(v') = 1$, and so that for any $v''$ with $S(v'') \subset S(v')$, $f(v'') = 0$.

\section{Results for Our Baseline Model}
\label{sec:opt_pess}

\subsection{Optimistic Approach}

In this subsection, we say an SCF $f$ is self-maintaining (omitting the qualifier ``for belief $F$'') if there \textit{exists} some common belief $F$ so that $f$ is self-maintaining for belief $F$. 

\begin{definition}
We fix an ordering of the voting vectors in $\Omega$: $v^1, \ldots, v^{|\Omega|}$ so that $v^{|\Omega| - 1} = \mathbf{0}, v^{|\Omega|} = \mathbf{1}$, and define the following distribution $F$:

$$Pr_{\hat{v} \sim F}[\hat{v} = v^i] = \begin{cases} \frac{2}{3} (\frac{1}{3})^{i-1} & i < |\Omega| \\ (\frac{1}{3})^{i-1})& i = |\Omega| . \end{cases}$$

We call $F$ ``lexicographic'' because it puts more weight on a voting profile $v^i$ than on all subsequent voting profiles $v^{i'}$ with $i' > i$ combined. 

\end{definition}

\begin{definition}[Vector negation and related properties.]
\label{def:proper}
 We say a vector $\neg v$ is the negation of vector $v$ if for any $1\leq i\leq n$, $\neg v_i = 1 - v_i$. 

We say that an SCF is \textit{proper} if there is no $v \in \Omega$ so that $f(v) = f(\neg v) = 1$. 

We say an SCF is \textit{decisive} if $\forall v \in \Omega, f(v) = 1 - f(\neg v)$. 

We say an SCF is 
    \textit{Null-preserving} if $f(\mathbf{0}) = 0$ (when all voters oppose a motion, is it rejected).
    
    We say an SCF is \textit{unanimity-preserving} if $f(\mathbf{0}) = 0, f(\mathbf{1}) = 1$. 
\end{definition}

\begin{restatable}{theorem}{OptArbitrary}
\label{thm:opt_arbitrary}
In the optimistic approach with worst-case tie-breaking and any domain $D_G, D_S, D_W$, any SCF $f$ is self-maintaining if and only if it is proper. In particular, since all SCFs in $D_W$ are proper, under the domain $D_W$ all SCFs are self-maintaining. 
\end{restatable}
\begin{proof}[Proof Sketch]
The intuition for why the lexicographic belief supports any proper voting rule with worst-case tie-breaking comes from single-issue voters. If the voter's belief is that a certain single decision is considered w.p. $1$, then if at the current voting rule it is expected to match their preference, and in another rule it is expected not to, the voter strictly supports maintaining the same rule, and vice versa. Thus, they vote as they would vote over the voting rule change as they vote over the decision itself if the decision is rejected under $f$, and (one can verify) opposite to it if the decision is accepted under $f$. The proper axiom handles the second case. 

However, single-issue voting is not a good choice of distribution because it has decisions with no probability support. Then, if we consider a change in the voting rule that only applies to non-supported decisions, the voters are indifferent to the change in voting rule, and worst-case tie-breaking can go any way. The \textit{lexicographic distribution} is useful in capturing both a behavior similar to single-issue voters, and full support. 
\end{proof}

We next consider the case of status-quo precedence (SQP). 
On the positive side, since worst-case tie-breaking is a stricter tie-breaking form than SQP, any SCF characterized as self-maintaining is also self-maintaining with SQP. However, it allows strictly more SCFs.

\begin{restatable}{theorem}{OptSQP}
\label{thm:opt_SQP}
In the optimistic approach with SQP tie-breaking, any SCF $f$ is self-maintaining if and only it is either proper or null-preserving. In particular, since all SCFs in $D_W$ and $D_S$ are null-preserving, for both domains the entire domain is self-maintaining. 
\end{restatable}

\subsection{Pessimistic Approach}

In this subsection, we say an SCF $f$ is self-maintaining if for \textit{any} common belief $F$, $f$ is self-maintaining for belief $F$. 

We start by considering worst-case tie-breaking.

\begin{restatable}{theorem}{PessArbitrary}
\label{thm:pess_arbitrary}
    In the pessimistic approach with worst-case tie-breaking and domain $D_G$, only the constant $0$ SCF is self-maintaining. In domains $D_S, D_W$, no SCF is self-maintaining. 
\end{restatable}

The intuition for this theorem is simple: If the distribution over possible decisions is not fully supported, there are voting rules that behave exactly the same as the incumbent voting rule. The voters are then indifferent to changing the rule, and by worst-case tie-breaking may vote in any way, including such that change the voting rule (unless $f = 0$, where \textit{no} vote changes the voting rule).

However, this argument seems restrictive. For example, for a dictator to give up their dictator power in $f^i_{dict}$ (which regardless of the distribution $F$ gives them the optimal utility of $1$), they must have very high trust in the new system of governance, and also favor change despite having perfect utility under $f^i_{dict}$. \footnote{Such examples are known in history. The Roman emperor Diocletian, who is known as the only emperor to voluntarily abdicate his title \cite{gibbon1833history}. Diocletian was sickly, and has been experimenting with delegating his authority throughout his rule
. It stands to reason that his belief was that foregoing power would preserve the stability of the decision-process he had built, and ``tie-breaking'' favored it due to his sickness. A similar dynamic is presented theatrically in Shakespeare's ``King Lear'' \cite{shakespeare2001king}, where the king abdicates his rule in favor of his three beloved daughters, due to his old age. In both cases (Lear and Diocletian), the beliefs driving the abdication decision turned out to be unsubstantiated.}

We thus consider our second tie-breaking rule, status-quo precedence. 
In Theorem~\ref{thm:pess_SQP_necessary} of the appendix, we provide a list of necessary conditions for an SCF to be pessimistic self-maintaining with SQP tie-breaking. For the domains $D_S$ and $D_W$, the necessary conditions provide a tight characterization. 

\begin{definition}[$k$-Oligarchy, Dictatorship, Anti-Dictatorship, Quasi-dictatorship]

    Let $K$ be a set of voters of size $k$. A voting rule that follows the strict majority vote among $K$ is called a $k$-Oligarchy. In particular, a $1$-oligarchy is a \textit{dictatorship}. 

    A voting rule $f$ that for some voter $i$ has $f(v) = 1 - v_i$ is called an \textit{anti-dictatorship}. 

    Let $S$ be a set of voter pairs so that some voter $i$ belongs to all pairs. Let $f$ be the monotone SCF where $S$ is the set of minimal winning coalitions. We say that $f$ is a \textit{quasi-dictatorship}. I.e., in a quasi-dictatorship, there is a quasi-dictator $i$ that has veto power, but needs to pair with other voters (from a given set) in order to pass decisions. 
\end{definition}

\begin{restatable}{lemma}{MinimalWinningTwo}
\label{lem:3olig_pdp}
    If $f$ is a proper monotone SCF where all the minimal winning coalitions are of size $2$ (pairs), then $f$ is either a $3$-oligarchy or a quasi-dictatorship. 

    If $f$ is a proper monotone SCF with a minimal coalition of size $1$, then it is a dictatorship. 
\end{restatable}

The lemma follows from a simple combinatorial analysis, and we prove it in the appendix. 

To provide intuition for our concluding theorem, consider an SCF where two voters that always get their way when they are in agreement. Then, as a pair, this SCF is optimal for them, in the sense that there is no SCF $f'$ they both strictly prefer. It can not benefit them when they are in agreement, so it can only benefit one on the expense of the other when they are in disagreement. Thus, if we also always need their mutual agreement to move away from $f$, we would never get it (under SQP tie-breaking). 

\begin{restatable}{theorem}{PessSQPChar}
\label{thm:pess_SQP_char}

A dictatorship, a $3$-oligarchy, and any quasi-dictatorship, are self-maintaining in the pessimistic approach with SQP tie-breaking. 

For the domains $D_S, D_W$, this is a complete characterization of the self-maintaining SCFs. 
\end{restatable}

In Example~\ref{ex:gap} of the appendix, we show a gap between our necessary and sufficient conditions for the general domain $D_G$, i.e., an SCF outside the class of Theorem~\ref{thm:pess_SQP_char} that is self-maintaining. 

\subsection{The I.I.D. Approach}
\label{sec:iid}

In this section, we consider any belief $F$ to be a joint distribution of an i.i.d. distribution over $\{0,1\}$. I.e., for some $F_p = Bernoulli(p)$, $F = F_p \times \ldots \times F_p$. If $1 > p > \frac{1}{2}$, i.e., the probability of each voter to prefer $1$ in a randomly drawn decision exceeds $\frac{1}{2}$, we say the voters are \textit{change-inclined}. If $p = \frac{1}{2}$, we say the voters are \textit{unbiased}. If $0 < p < \frac{1}{2}$, we say the voters are \textit{change-averse}. We ignore the degenerate cases $p = 0$ and $p = 1$. 


\begin{restatable}{theorem}{UnbiasedVoters}
\label{thm:unbiased_voters_char}
For \textit{unbiased} voters, an SCF $f$ is self-maintaining with worst-case tie-breaking for the domain $D_G$ if and only if it is either a dictatorship, anti-dictatorship, or the constant $0$, and for the domain $D_S$ if and only if it is a dictatorship. 
\end{restatable}

The proof for the simple games domain $D_S$ is short and we include it here. The proof for $D_G$ is more involved, as it can not make use of minimal winning coalition, and it appears in the appendix. 

\begin{proof}[Proof for $D_S$]
    Let $S$ be the largest-size minimal winning coalition. If $S$ is of size $1$, and $f$ is not a dictatorship, let the single voter in $S$ be denoted $i$. If we take $f'$ to be the $i$-dictatorship, we have $c_i(f,f') = 1$, and since $i$ can then pass the decision, $f$ is not self-maintaining. 
        
        If $|S| \geq 2$, fix some two voters $i, j \in S$. Take a maximal losing coalition that contains $S \setminus \{i\}$ (such maximal losing coalition must exist since $S \setminus \{i\}$ is a losing coalition). Denote it $L_i$, and similarly take a maximal losing coalition that contains $S \setminus \{j\}$ and denote it $L_j$. Let $f'$ be the SCF that has all the winning coalitions of $f$ as well as $L_i, L_j$. $f'$ is monotone since we only added maximal losing coalitions into the winning coalitions, and is unanimity-preserving since $L_i, L_j$ are each of size at least $1$ (and so $f'(\mathbf{0}) = f(\mathbf{0}) = 0$). 
All voters $S\setminus \{i,j\}$ strictly prefer $f'$, while $i,j$ have the same utility for $f, f'$. We can thus choose $\mathbf{c}(f,f')$ so that for any $i' \in S$, $\mathbf{c}_{i'}(f,f') = 1$. Since $S$ is a minimal winning coalition for $f$, this results in a violation of self-maintenance. 
\qed \end{proof}

The domain of weighted majority games does not admit the above characterization. We show a general example for the gap.

\begin{example}
\label{ex:maximal_quasi_dictatorship}
    For $n\geq 3$, consider $f$ to be the quasi-dictatorship where the minimal winning coalitions are some voter $i$ and any other voter $j$ (call it the \textit{maximal quasi-dictatorship}). It can be implemented as a weighted majority game using the weights: $w_i = n$, and $w_j = 1$ for any $j\neq i$, and threshold $T = n$. For the unbiased i.i.d. distribution, we have $u_i(f, F) = 1 - \frac{1}{2^n}$, as the only voting preference where voter $i$ does not agree with the SCF outcome is when $v_i = 1$ and all other voters have $v_j = 0$. For any other voter $j$, we have $u_j(f, F) = Pr[v_j = 0] \cdot Pr[f(\mathbf{v}) = 0 | v_j = 0] + Pr[v_j = 1] \cdot Pr[f(\mathbf{v}) = 1 | v_j = 1] = \frac{1}{2}\left( (Pr[v_i = 0] + Pr[v_i = 1, \forall j' \not\in \{i,j\} v_{j'} = 0]) + Pr[v_i = 1] \right) = \frac{1}{2}\left(\frac{1}{2} + \frac{1}{2^{n-1}} + \frac{1}{2} \right) = \frac{1}{2} + \frac{1}{2^n}$. 

    Since $i$ is a veto player for $f$, we must have $c_i(f,f') = 1$ for the self-maintaining condition to be violated. Thus, we only need to consider $f'$ so that $u_i(f', F) \geq u_i(f, F)$. Since there is exactly one voting preference $\mathbf{v}$ where $f$ does not match voter $i$'s preference (namely, where $v_i = 1$ and all other voters have $v_j = 0$), any such $f' \neq f$ must have $f'(\mathbf{v}) = 1$. Since $f'$ is a weighted majority game SCF, it is monotone, and so $f'$ must agree with voter $i$ whenever it votes $1$. Moreover, as a weighted majority game SCF it is also proper, and so $f'(\neg \mathbf{v}) = 0$. By monotonicity, this means $f'$ must agree with voter $i$ whenever it votes $0$. All in all we conclude that $f'$ must be the $i$-dictatorship. However, the utility of all other voters $j$ in the $i$-dictatorship is exactly $\frac{1}{2} < \frac{1}{2} + \frac{1}{2^n} = u_j(f, F)$, and so $c_i(f, f') = 1$, but for all other voters $j$, $c_j(f, f') = 0$. This yields $f(\mathbf{c}(f,f')) = 0$, and $f$ is self-maintaining. 
\end{example}



\section{Extensions and Discussion}
\label{sec:discussion}

\subsection{Forward-Looking Voters}
\label{sec:forward_looking}

Our formalization of stability takes a $1$-step-ahead perspective, where an SCF is self-maintaining if and only if there is no coalition that prefers another SCF. However, it is possible that coalitions (and in particular, members of coalitions) take a more sophisticated approach: They understand that while $f'$ is preferable over $f$, it might then lead to $f''$, which is worse than $f$. Thus, they prefer to not move to $f'$, and this makes $f$ self-maintaining. We formalize this forward-looking approach and demonstrate it using the unbiased i.i.d. case, showing that indeed it strictly extends the set of self-maintaining SCFs. 

Fix some belief $F$, and a monotone domain $D_M$. We consider monotone domains since this guarantees that all SCFs are monotone, and we can define transitions using a coalition $S$ independent of how voters outside $S$ behave. 
We define forward-looking self-maintaining SCFs. 
Our definition fits within the framework of Chwe \cite{chwe1994} for the farsighted core. 

\begin{definition}
    For two SCFs $f, f' \in D_M$ We say that $f \rightarrow_S f'$ if there exists a  coalition $S$ of $f$ so that $\forall i\in S, u_i(f', F) > u_i(f, F)$, and $f(v(S)) = 1$. 
    This defines a directed graph $(D_M, E)$, where the edges are the SCF links. 
    For a coalition $S$ so that there is a path from $f$ to $g$ that starts with $f \rightarrow_S f'$ for some $f'$, and also $\forall i \in S, u_i(g, F) > u_i(f,F)$, we say that $f \Longrightarrow_S g$. 
    
    We say a set of outcomes $Y$ is a \textit{stable set} of the forward-looking self-maintaining problem if the following conditions hold:

    \begin{enumerate}

        \item (Internal consistency) $\forall f, g \in Y$, for any coalition $S$, it does not hold that $f \Longrightarrow_S g$. 

        \item (External consistency) $\forall f \not \in Y, \exists g \in Y$ and a coalition $S$ so that $f \Longrightarrow_S g$. 
 
    \end{enumerate}

The core of the forward-looking self-maintaining problem is the intersection of all stable sets. For an outcome in the core we say it is forward-looking self-maintaining w.r.t. $F, D_M$. 
    
\end{definition}



\begin{lemma}
    Any SCF $f$ that is self-maintaining with SQP tie-breaking for a belief $F$ and domain $D_M$ is in any stable set of the corresponding forward-looking self-maintaining problem. 

    Thus, if there exists a stable set of the forward-looking self-maintaining problem, then $f$ is forward-looking self-maintaining w.r.t. $F, D_M$. 
\end{lemma}

\begin{proof}
    Since $f$ is self-maintaining with SQP tie-breaking, there is no such $f'$ so that a coalition of $f$ strictly prefers moving to $f'$. Since our definition of $f \Longrightarrow_S g$ requires that $f \rightarrow_S f'$ for some $f'$ on the first edge out of $f$ on the path to $g$, by external consistency, $f$ must belong to any stable set. 
\qed \end{proof}


We consider the following process to identify forward-looking self-maintaining SCFs beyond these that are self-maintaining with SQP tie-breaking. Consider $f''$ that is self-maintaining with SQP tie-breaking, and $f'$ so that there is a minimal winning coalition $S$ that strictly prefers $f''$ over $f'$. Then, since $f' \rightarrow_S f''$, it holds that $f' \Longrightarrow_S f''$. Since $f''$ is in every stable set, $f'$ can not be in any stable set due to internal consistency, and thus can also not appear in the core. 
We call such $f'$ a ``transient'' SCF. 

Now consider an SCF $f$ so that for any $f'$ with $f \rightarrow_S f'$, it holds that $f'$ is transient. Consider if there is a stable set $Y$ so that $f \not \in Y$. By external consistency, there must be some $g\in Y$ and a coalition $S$ so that $f \Longrightarrow_S g$, and there must be a transient $f'$ that is part of this path. Thus, it also holds that $f' \Longrightarrow_{S'} g$ for some coalition $S'$. We conclude that for $f$ to \textit{not} appear in some stable set, there must be a coalition $S'$ with $f'(v(S')) = 1$ and a coalition $S$ with $f(v(S)) = 1$ that agree $g$ is better than the (respective) incumbent SCF. If there is \textit{no} such coincidence for $f$ and \textit{any} of the transient SCFs $f'$ it leads to, we conclude that $f$ must be in every stable set. It is thus in the core and is forward-looking self-maintaining w.r.t. $F, D_M$. 

This process (which is a form of an iterated core, see \cite{chwe1994,gillies1959solutions}), can repeat until the ``transient'' and ``forward-looking self-maintaining'' sets stabilize. I.e., we start with the self-maintaining with SQP tie-breaking set as the initial forward-looking self-maintaining set, and an empty transient set. At each step, we derive an updated transient set based on the current forward-looking self-maintaining set, and then an updated forward-looking self-maintaining set based on the transient set. We continue until the sets do not change. By our argument, any SCF in the forward-looking self-maintaining set is guaranteed to be so. A transient SCF is necessarily \textit{not} forward-looking self-maintaining in any solution of the game. 

We run this procedure for $3 \leq n \leq 5$ on the domain $D_W$ of Weighted Majority Games, and $F$ the unbiased i.i.d. distribution. 

There are $11$ such SCFs for $n=3$, $80$ such SCFs for $n=4$, and $1683$ for $n=5$. We provide the full details of our computational results in the appendix. Importantly, we find that for $n=4$ and $n=5$, there are forward-looking self-maintaining SCFs that are not self-maintaining with SQP tie-breaking, have better Nash welfare, and weakly better social welfare (in the $n=5$ case, strictly better). 

We also note the following: 
If the process terminates where it identifies \textit{all} states as either forward-looking self-maintaining or transient, then we have fully characterized the forward-looking core. This happens in our $n=4$ run. 
If there are unidentified states, then we can further attempt a complementary procedure: We directly encode the internal and external consistency conditions as a SAT problem, with an optimization goal of finding a minimal stable set. Then, for $100$ times we repeat a random SAT run (with no optimization goal) to find other stable sets. Since every forward-looking self-maintaining SCF must be in the intersection of all these random attempts, they can help rule out SCFs that are outside the core. For our $n=5$ run, we find out that the set of SCFs identified as forward-looking self-maintaining by the iterated core process is conclusive, and no other SCFs are forward-looking self-maintaining. 

Therefore, in both $n=4,5$ cases, the iterated process found all forward-looking self-maintaining SCFs. 


\subsection{Bi-Level Voting}
\label{sec:bilevel}

We consider a generalization of our framework, where the aggregating SCF of the first stage $\hat{f}$ is not necessarily equal to the incumbent rule $f$ of the second stage. 
In this bi-level generalization, we revise Definition~\ref{def:self_maintaining} as follows:

\begin{definition}
    We say a bi-level constitution $f, \hat{f} \in D$ is self-maintaining for belief $F$ and domain $D$ if for any other SCF $f' \in D$, and any choice $\mathbf{c}$ that is compatible with voter utilities toward $f, f', F$ and the tie-breaking rule we consider, 

    $$\hat{f}(\mathbf{c}(f,f',F)) = 0.$$
\end{definition}

While this generalization is relevant to all our approaches, domains, and tie-breaking rules,
in this subsection we focus on the case we find most interesting, which is the pessimistic approach with SQP tie-breaking for Weighted Majority Games. 


With $\hat{f} = Unanimity$, there is an equivalence between self-maintenance and a form of Pareto-Efficiency of the second-stage SCF $f$.

\begin{definition}
    \textit{Pessimistic Pareto-Efficient.} We say an SCF $f$ is Pessimistic Pareto-Efficient for domain $D$ if for any distribution $F$, there is no $f'\in D$ so that $\forall i, u_i(f',F) > u_i(f, F)$. 
\end{definition}

\begin{restatable}{theorem}{PessimisticParetoEfficient}
\label{thm:pess_pareto}
With $\hat{f} = Unanimity$, a bi-level constitution $f, \hat{f}$ is self-maintaining with the pessimistic approach and SQP tie-breaking if and only if it is pessimistic Pareto efficient. 
\end{restatable}

We provide some necessary and sufficient conditions beyond this general characterization. 

\begin{restatable}{proposition}{ThreePartition}
\label{prop:three_partition}
    Let $S_1, S_2, S_3$ be a $3$-partition of the voters $[n]$. Let $v_{i,j}$ be the decision where $S_i, S_j$ vote $1$ and the rest vote $0$. If $\forall 1 \leq i \neq j \leq 3$, $f(v_{i,j}) = 0$, then for any simple game $\hat{f}$, it holds that the bi-level constitution $f, \hat{f}$ is not pessimistic self-maintaining with SQP tie-breaking. 
\end{restatable}

\begin{restatable}{corollary}{WMGThresh}
Consider a WMG $f$ with total weights $W = \sum_{i=1}^n w_i$, threshold $T$, and maximal individual weight $\mu = \max_{i=1}^n w_i$. If $\mu < \frac{T}{2} - \frac{W}{3}$, then for any WMG $\hat{f}$, it holds that the bi-level constitution $f, \hat{f}$ is not pessimistic self-maintaining with SQP tie-breaking. 
\end{restatable}

\begin{restatable}{proposition}{DecisiveConstitution}
    If a WMG is decisive (Definition~\ref{def:proper}),  then the bi-level constitution $f, Unanimity$ is pessimistic self-maintaining with SQP tie-breaking.
\end{restatable}

In particular, this holds for any oligarchy with an odd number of voters.


\subsection{Conclusion and Future Directions}

Our work takes a fresh look at the idea of self-maintaining SCFs, with a novel focus on stability within different domains, and on applying the rationalizability and robustness notions to self-maintenance. However, our results, especially in the pessimistic and i.i.d. approach are mostly negative. In the perspective of a designer that wishes to enable good SCFs (in terms of Nash and social welfare), this presents a problem. We examine two analytic approaches that might uncover more self-maintaining SCFs (forward-looking voters in Section~\ref{sec:forward_looking}, and incorporating common value in Appendix~\ref{sec:jury}), as well as the bi-level design pattern (in Section~\ref{sec:bilevel}). There are other important directions to try: (i) While the forward-looking approach shows promise in our analysis of $n=4$ and $n=5$, we do not provide a general $n$ separation, i.e., a construction for every $n$ of an SCF which is forward-looking self-maintaining but not self-maintaining with SQP tie-breaking, or a counter-example that shows such construction is not possible. 
(ii) Abstentions are common in practice and may effect our analysis. Generalizing our framework to non-binary decisions is also interesting, although in social choice the binary case is usually the ``better-behaving'' case (e.g., Gibbard-Satterthwaite's negative conclusion applies only when there are more than two alternatives). (iii) More generally, a unique property of DAOs is that they run on top of smart contracts that are Turing-Complete. Thus, we can think of SCFs that have a very general form, much more general than matching rankings to a decision as it is commonly treated in social choice. Whether there is a way to create meaningful self-maintaining rules in this way is very interesting. In particular, it could be possible to handcraft the domains of legible SCFs (rather than consider entire natural classes such as $D_G, D_S, D_W$ as we do), creating domains where good SCFs emerge as self-maintaining.

\bibliographystyle{splncs04}
\bibliography{refer.bib}

\appendix

\section{LLM Usage Disclosure.}

The initial python code for the equilibrium computations of Section~\ref{sec:forward_looking} and the simulations of Appendix~\ref{sec:jury}, including the code that generated the figures, was developed in cooperation with Gemini $3$ Pro. Gemini $3$ Pro was also essential in uncovering related work, most notably the line of work on Stable Constitutions, and US and EU legal precedents. The work has benefited from comments by the AI feedback tool Refine.ink, and from issues surfaced while auto-formalizing the manuscript using Aristotle \cite{aristotle}. In particular, the entire statement and proof of Theorem~\ref{thm:anonymous_opt}, as well as corrections to the proof of Theorem~\ref{thm:iidBiased} and Lemma~\ref{lem:nested_olig_char}, are due to Aristotle. The Lean4 formalization of the paper is available at \href{https://github.com/yotam-gafni/centralization_stability_formal_constitutions}{https://github.com/yotam-gafni/centralization\_stability\_formal\_constitutions}.

\section{Additional Related Work}

We are inspired by very early political science work, going back to Plato, Aristotle, and Polybius \cite{lord2013aristotle,reeve2004plato,mcging2010polybius}. Plato considered how all systems of governance end up in tyranny. Aristotle considered, based on different case-studies of his time, how government forms may change, without concluding in some inevitable end-state. Polybius presented a cyclical theory of constitution where dictatorship transforms to oligarchy, then to democracy, and then back to dictatorship. One interesting conclusion by Polybius is that to arrive at a stable constitution, one must mix elements of the different government forms. 
These type of dynamics were on display throughout history:
In the French revolution \cite{doyle2018oxford}, the monarchy was replaced by a full democracy, but then later Robespierre's ``Committee of Public Safety'' (which may be thought of as a small oligarchy) assumed greater powers, and eventually the power struggle ended up with Napoleon as emperor (i.e., a return to monarchy). During the Communist revolution in Russia \cite{brovkin1987mensheviks}, the two factions of Bolsheviks and Mensheviks both fought for Communist rule against the White Army, but then later the Menshevik party was banned. 

Another illustrative example is that of Unanimity, the voting rule where all voters must support a motion for it to pass. 
Voters prefer to move away from Unanimity, and risk losing sometimes, over having a deadlocked system. This is supported by significant precedents. In the US, the Articles of Confederation \cite{articlesConfederation} required all $13$ states to support a motion for it to pass. However, it was soon superseded by the US constitution \cite{USconstitution}, which only requires a qualified majority of $9$ out of $13$ states. More recently, Article $48$ of the Treaty on European Union \cite{teu_art48} originally required decision by Unanimity by the European Council. However, the Treaty of Lisbon \cite{treaty_lisbon_art48_2007} moved many policy areas to be decided by a qualified majority. Here is an emblematic quote, from the classic textbook on EU Law by Craig \& de Búrca \cite{craig2011eu}, page 133: ``Whether Unanimity is the best protector of national sovereignty depends therefore on whether a state believes that maximizing the possibility of inaction [...] is better for the national interest than a qualified-majority voting rule which increases the possibility of action''. They give a concrete example from the UK: ``The Unanimity rule [...] impeded the market liberalization desired by the Conservative Party, [...] hence the willingness to sacrifice the veto for the enhanced possibility of Community action''. 

We also note that established modern work on civil resistance finds that the most important factor of successful protest is in the ability to shift the loyalties of regime insiders \cite{civilResistance}. This paints a picture where ``revolutions'', while seemingly an outside force, succeed mostly through realigning the current power structure, in accordance with our De-Jure approach. 

\subsubsection{More Connections to Stable Constitutions Work}

Works following Koray \cite{koray2000self} attempted (and sometimes succeeded) to circumvent his Arrow-style impossibility by imposing certain restrictions. We are most interested in restricting the \textit{domain} of SCFs, as we take a similar approach. 
Ünel \cite{unel1999} considers SCFs that only use the top priority in each voter's preference, but arrives at the same dictatorship characterization as Koray \cite{koray2000self}. Slinko \& Koray \cite{slinkoKoray2008} only allow Pareto-Optimal alternatives to be chosen in any given preference profile, which allows for a meaningful Pareto-Optimal anti-dictatorship rule to emerge. However, restricting to Pareto-Optimal alternatives is less meaningful in the binary case (it is akin to respecting Unanimity, which all Simple Games do). Altuntaş \cite{Altuntas} notes that every SCF is self-selective in \textit{some} domain (as it is at least self-selective in the singleton domain that only includes itself), and uses it to define degrees of self-selectivity. Concretely, it considers a family it defines of $q$-Condorcet rules, and shows that rules with lower $q$ values are self-selective w.r.t. rules with higher $q$ values. It also shows a rule called Majoritarian Compromise which is self-selective w.r.t. any $q$-Condorcet rule, when there are $n=3$ voters. Kerman \& Koray \cite{kerman2023nested} show that self-selectivity is closed under intersection and union of sets: If $f$ is self-selective in a domain $D_1$ and $D_2$, then it is also self-selective in $D_1 \cap D_2$ and $D_1 \cup D_2$ (in our setting, this is an immediate property of self-maintenance under the pessimistic and i.i.d. approaches and tie-breaking rules, and holds in the optimistic approach as well due to our generic distribution construction). 

Jeong \& Kim \cite{jeongKim2023,JeongKim2024} consider respectively the interim and ex-post stability case, where bidders know their own preference for the vote, but not others (interim), or know everyone's preference (ex-post). This is different from our ex-ante approach. We find the ex-ante case most natural when facing unknown future decisions. However, on a technical level, the ex-post characterization of self-stability by Jeong \& Kim \cite{JeongKim2024} as all \textit{proper} SCFs (Corollary~2) is similar with our optimistic approach result, and for a good reason. Both rely to some extent on voters being ``Single Issue Voters''. See the intuition we provide for our proof.

\subsubsection{Models for Expansion and Retraction of Voting Rights.}

Roberts \cite{roberts1999}, as well as Jack \& Lagunoff \cite{JackLagunoff2006} consider club membership models where current club members may extend membership to new members. This is similar to our model of nested oligarchies in Appendix~\ref{sec:jury}.

Hermida-Rivera \cite{stableSimpleGames} considers stability of voting rules when voters aim to maximize their power indices in simple games. When applied to WMGs, their results greatly simplify and shows that (with their additional axioms) only dictatorships are minimally stable. 
Our \textit{extractive} component corresponds to the same power-index utility model, but we also incorporate 
``Condorcet Jury Theorem'' effects, which allows for multiple stable rules, including a large committee. 

More broadly, the works on \textit{Institution Design} are highly relevant to us, most prominently Acemoglu \& Robinson \cite{acemoglu2001theory}. The driver of extending voting right in these works usually comes from either the threat external conflict (a revolution by non-voters), or by internal conflict (among the voters).  Our model suggests a (partial) democratization force we believe is novel, based on the quality of decisions:

The Condorcet Jury Theorem \cite{condorcet1785} is often cited in support of democracy as the optimal mechanism for ``wisdom of the crowd''. The probabilistic argument shows that if voters have i.i.d. access to signals regarding the best decision, with $p > \frac{1}{2}$ accuracy, then the best aggregation is taking a majority over each of the voters' signal. 
We make a nuanced application of it to SCF stability in Appendix~\ref{sec:jury}: We show that this serves as a force to have more inclusive stable SCFs. 

\subsubsection{DAO Governance and Centralization}

Research on Blockchain Decentralized Autonomous Organization (DAOs) takes different perspectives. One perspective is their effect of the larger blockchain ecosystem: How they may be used to facilitate strategic attacks (``Dark DAOs'') \cite{fabrega2025voting}, affect bidding patterns, due to their group nature \cite{bahraniDAOs}, how to protect minority opinion to maintain user retention \cite{gafniFairDM}, or quantifying the possible ``Anscombe Paradox'' effects of following the majority in many sequential decisions \cite{fritsch2022price}. 

The focus that is most relevant to our study is that of \textit{centralization} and security of DAOs \cite{balietti2025slayingdragonquestdemocracy,sokDAOs}. Centralization is generally a major concern in Blockchains \cite{centralizationBahrani,chemaya2025quantifying,levy2025analyzing}. 
A string of works \cite{whoControlsDAOs,fabrega2025voting,yaish2026inequality} define and study centrality measures for DAOs, and generally find that DAOs are more centralized than what we would like to believe. Falk et. al \cite{falk2024blockchain} find that in four major DAOs (Uniswap, Lido, Aave and Compound), if the top $3$ or $4$ token holders are in consensus, they could together decide almost every vote. They name this metric \textit{minimal quorum}, and it is also commonly known as the Nakamoto coefficient \cite{whoControlsDAOs,sokDecentralization}.
Kitzler et. al \cite{govDAO} find that in over $7.5\%$ of DAOs, the contributors (who develop and maintain the DAO) have majority control, and have solely dictated a decision in over $21\%$ of DAOs. Moreover, they find that token power balance (i.e., the voting structure) has shifted shortly before a decision in over $14\%$ of motions. Similarly, Cong et. al \cite{cong2025centralized} find abnormal trading activity in tokens around governance proposals. Barbereau et. al \cite{barbereau2023decentralised} find that not only are DAOs centralized, but are becoming more centralized with time. Our work offers a social-choice micro-foundation for all these phenomenons. 

\section{Application to Expansion and Retraction of Voting Rights}
\label{sec:jury}

In this extension, we incorporate our notion of self-maintenance within a certain domain to gain insight into expansion and retraction of voting rights. We consider a model where the class of voting rules are a nested sequence of oligarchies $O_1, \ldots, O_n$. An oligarchy $O_i$ is composed of a set of $S_i$ voters with $|S_i| = i$, and where $S_{i-1} \subset S_i$. The vote is decided by a majority of the set $S_i$, where a motion needs a strict majority to pass. This model allows to study questions such as whether a dictatorship ($O_1$) would prefer to open up to an oligarchy ($O_i$ with $1 < i < n$) or a full democracy ($O_n$), and vice versa. 
We now specify the voters' utilities, which is the main point of divergence from the model presented in Section~\ref{sec:model}, other than the domain change. The voter utility has two components, \textit{extractive} $E$ and \textit{participative} $P$, linearly weighted by a parameter $0 \leq \lambda 1$, so that $u_j(O_i) = \lambda \cdot E_j(O_i) + (1-\lambda) \cdot P_j(O_i)$. The extractive component depends on whether the voter is part of the oligarchy, and if it is, it gets its relative share:
$$E_j(O_i) = \frac{1}{i} \cdot 1[j \leq i].$$
For the participative part, we draw from the Condorcet Jury model. We consider that there is a true state $s$, and each voter has an i.i.d. signal with $p >  \frac{1}{2}$ w.r.t. it. This means the overall number of successful prediction $k$ is drawn from the binomial distribution with parameter $p$ (formally, $Binomial(i,p)$ is the binomial distribution with $i$ attempts and success probability $p$). 

Thus, for any $j$, 
$$P_j(O_i) = Pr_{k\sim Binomial(i,p)}[k > \frac{i}{2}], $$

is the probability that the right decision is made under oligarchy $O_i$. The pessimistic self-maintenance with SQP tie-breaking condition for oligarchy $i$ given parameters $\lambda, p$ is that for any $i' \neq i$, a (weak) majority of the members of $O_i$ have $u_j(O_i) \geq u_j(O_{i'})$. For the rest of this section, we use self-maintaining in short to mean pessimistic self-maintaining with SQP tie-breaking. 

We make a few observations. With $\lambda = 1$ (a fully extractive economy), and $i\geq 3$, if we let $i' = \lceil \frac{i+1}{2} \rceil$, then all voters $j \leq i'$ prefer $O_{i'}$ over $O_i$, and they constitute a strict majority in $O_i$. Thus, the only self-maintaining oligarchies are $O_1$ (a dictator) and $O_2$ (a quasi-dictatorship). With $\lambda = 0$ (a fully participative economy), the probability that the majority of a Binomial distribution matches the true state is monotone increasing in the odd numbers, and therefore the only self-maintaining rule is the full democracy $O_n$ (if $n$ is odd) or $O_{n-1}$ if $n$ is even. Aside from the illuminating edge cases, for the rest of the section we consider the case of $0 < \lambda < 1$, that mixes extraction and participation. We also consider a large populace (an infinite $n$, though the results naturally apply to any $n$ that is at least as large as the largest stable outcome). 

To explicitly calculate the full set of self-maintaining oligarchies for given values of $\lambda, p$, we do the following computational process. 
We start with a necessary condition for self-maintenance, based on the idea we applied with $\lambda = 1$. I.e., we require 
$u_{i'}(O_i) \geq u_{i'}(O_{i'})$ for the voter $i' = i - 1$ (with $i\geq 3$). Note that all $j < i'$ have $u_j(O_i) = u_{i'}(O_i), u_j(O_{i'}) = u_{i'}(O_{i'})$, and so if the reverse holds (and $u_{i'}(O_i) > u_{i'}(O_{i'})$, then we have a violation of self-maintenance as in the $\lambda = 1$ case. 
We get:

\[
\begin{split}
& \frac{\lambda}{i} + (1-\lambda) \geq \frac{\lambda}{i} + (1-\lambda)Pr_{k\sim Binomial(i,p)}[k > \frac{i}{2}] \\
& \geq \frac{\lambda}{i'} + (1-\lambda)Pr_{k\sim Binomial(i', p)}[k > \frac{i'}{2}] \stackrel{Hoeffding}{\geq} \frac{\lambda}{i'} + (1-\lambda)(1 - e^{-2i'(p-0.5)^2}).
\end{split}
\]

It turns out this simple bound for the necessary condition is enough to filter out large values of $i$. With the remaining candidates for self-maintaining oligarchies, we require:
\begin{definition}
    \textit{Right-Maximal.} For a function $f$ and a value $i^*$, we say it is right-maximal if $\forall i' > i^*$, $u_{i^*}(O_{i^*}) \geq u_{i^*}(O_{i'})$. 

    \textit{Halving-Safe.} For a function $f$ and a value $i^*$, we say it is halving-safe if $\forall \frac{i^*}{2} < i' < i^*$, $u_{i'}(O_{i^*}) \geq u_{i'}(O_{i'})$. 
\end{definition}

\begin{lemma}
\label{lem:nested_olig_char}
    For $i^*\geq 3$, an oligarchy $O_{i^*}$ is self-maintaining if and only if it is right-maximal and halving-safe. In particular, no $O_{i^*}$ with even $i^* \geq 4$ is self-maintaining.  
\end{lemma}

\begin{proof}
    (Both conditions are necessary)

    If right-maximal is violated, then there is $i'$ so that $u_{i^*}(O_{i'}) > u_{i^*}(O_{i^*})$. Since for all $i < i^*$, $u_i(O_{i^*}) = u_{i^*}(O_{i^*})$ and $u_i(O_{i'}) = u_{i^*}(O_{i'})$, this means all voters in $O_{i^*}$ would vote to move to $O_{i'}$ and $O_{i^*}$ is not self-maintaining.

    If halving-safe is violated, then there is $i' > \frac{i^*}{2}$ so that $u_{i'}(O_{i'}) > u_{i'}(O_{i^*})$. Since for all $i < i'$, $u_i(O_{i^*}) = u_{i'}(O_{i^*})$ and $u_i(O_{i'}) = u_{i'}(O_{i'})$, this means more than half the voters in $O_{i^*}$ (namely, these with $i\leq i'$) would vote to move to $O_{i'}$ and $O_{i^*}$ is not self-maintaining.

    (Satisfying both conditions is sufficient)

    If right-maximal is satisfied, then for any $i' > i^*$, $u_{i^*}(O_{i^*}) \geq u_{i^*}(O_{i'})$. Since for any $i\leq i^*$ we have $u_{i}(O_{i^*}) = u_{i^*}(O_{i^*})$, and $u_i(O_{i'}) = u_{i^*}(O_{i'})$, it holds for any $i\leq i^*$ that $u_i(O_{i^*}) \geq u_i(O_{i'})$, and so there is a consensus not to move away from $O_{i^*}$ to $O_{i'}$. 

    If halving-safe is satisfied, then for any $\frac{i^*}{2} < i' < i^*$, $u_{i'}(O_{i^*}) \geq u_{i'}(O_{i'})$, and so similarly for all $i\leq i'$, which constitute a majority of the voters in $O_{i^*}$, $u_i(O_{i^*}) \geq u_i(O_{i'})$, and there is a majority that does not wish to move away from $O_{i^*}$ to $O_{i'}$. 

    \textit{For an odd $i^*$}, for any $i' \leq \frac{i^*}{2}$, it holds for any $i' < i \leq i^*$ that $u_i(O_{i'})$ does not get any extractive utility, and its participative utility is lower than $O_{i^*}$ (which includes more voters). So, $u_i(O_{i'}) \leq u_i(O_{i^*})$. This means that there is no strict majority among voters in $O_{i^*}$ to move away from $O_{i^*}$ to $O_{i'}$. 

    \textit{For an even $i^* \geq 4$}, for $i' = i^* - 1$ it holds that $i' > \frac{i^*}{2}$, its participative utility is higher, and its extractive utility is higher for all $i \leq i'$. Thus, there is a majority supporting moving to $O_{i'}$, and $O_{i^*}$ is not halving-safe (and not self-maintaining). 
    
\qed \end{proof}

The lemma then allows us to precisely characterize the outcomes for given parameters: We consider whether $O_1$ and $O_2$ are self-maintaining, and for all $i^*\geq 3$ check whether $O_{i^*}$ is right-maximal and halving-safe. The results are summarized in the figures below:

\begin{figure}[htbp]
    \centering
    \includegraphics[width=0.6\textwidth]{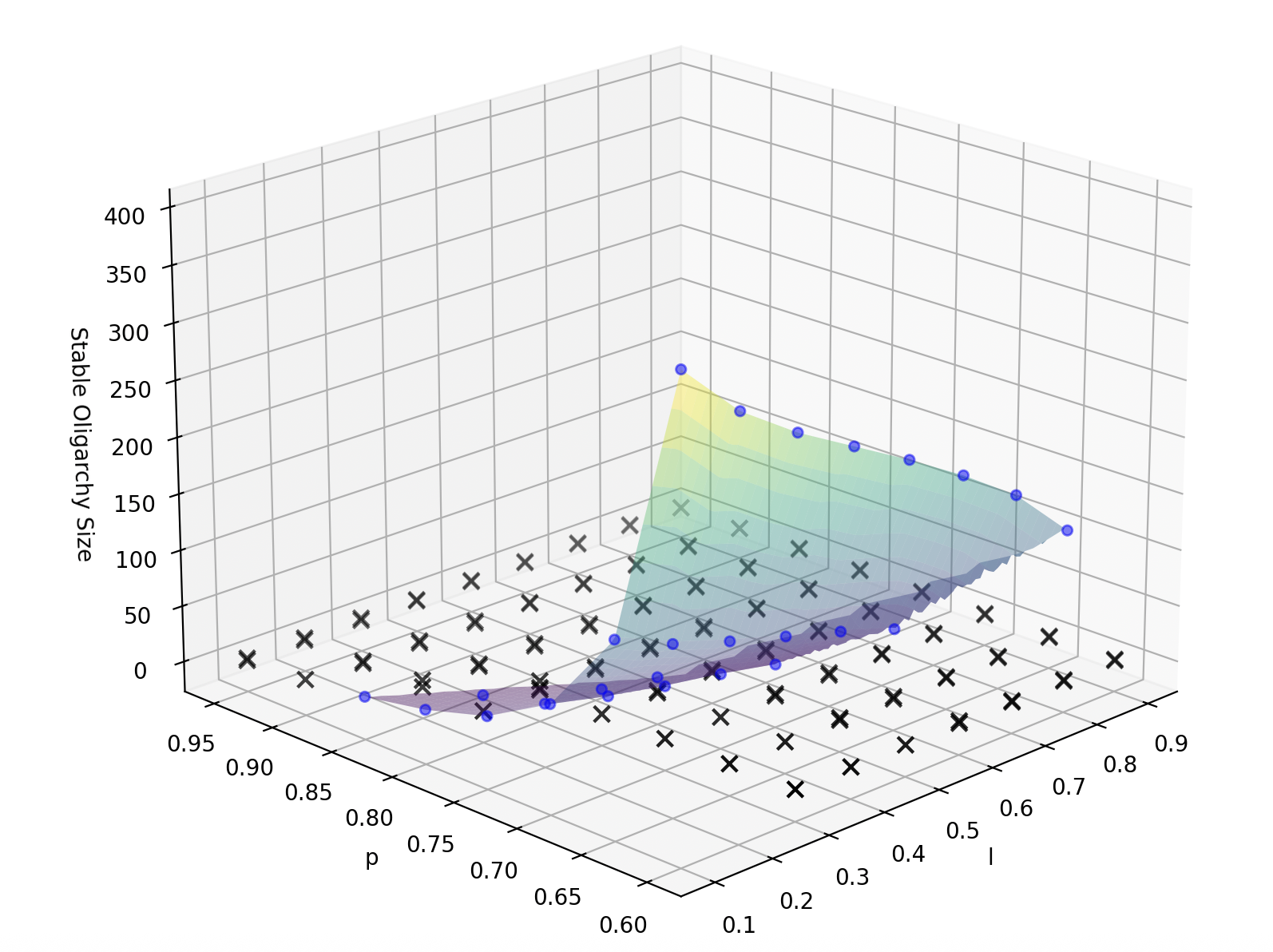}
    \caption{Stable oligarchies as a function of $\lambda$ and $p$. Black $x$ marks are used for small oligarchies (such as dictatorship, $2$-oligarchy and $3$-oligarchy), and blue circles for larger-value oligarchies (upwards of $i=10$).}
    \label{fig:3d_landscape}
\end{figure}


We run a Cartesian grid of parameters, with $\lambda \in \{0.1, 0.2, \ldots, 0.9\}$ and $p\in \{0.6,0.65, \ldots, 0.95\}$. 
Let us describe the results. At the corner of low $\lambda$ and low $p$ values, we see that \textit{the unique} solution is a large oligarchy (let us call it a committee). The size of committee is monotone-decreasing in both $\lambda$ and $p$. Then, when we move away from the low-values corner, we get some parameter values where there are multiple self-maintaining outcomes: A large committee, a dictatorship, and also possibly a small oligarchy (with $i\in \{2,3\}$). When $\lambda$ and $p$ values are high enough, large committees are no longer self-maintaining, and we remain only with the dictatorship and (possibly) a small oligarchy. 

\begin{figure}[htbp]
    \centering
    \subfloat[Stable oligarchies as a function of $\lambda$ with $p=0.6$]{
        \centering
        \includegraphics[width=0.48\textwidth]{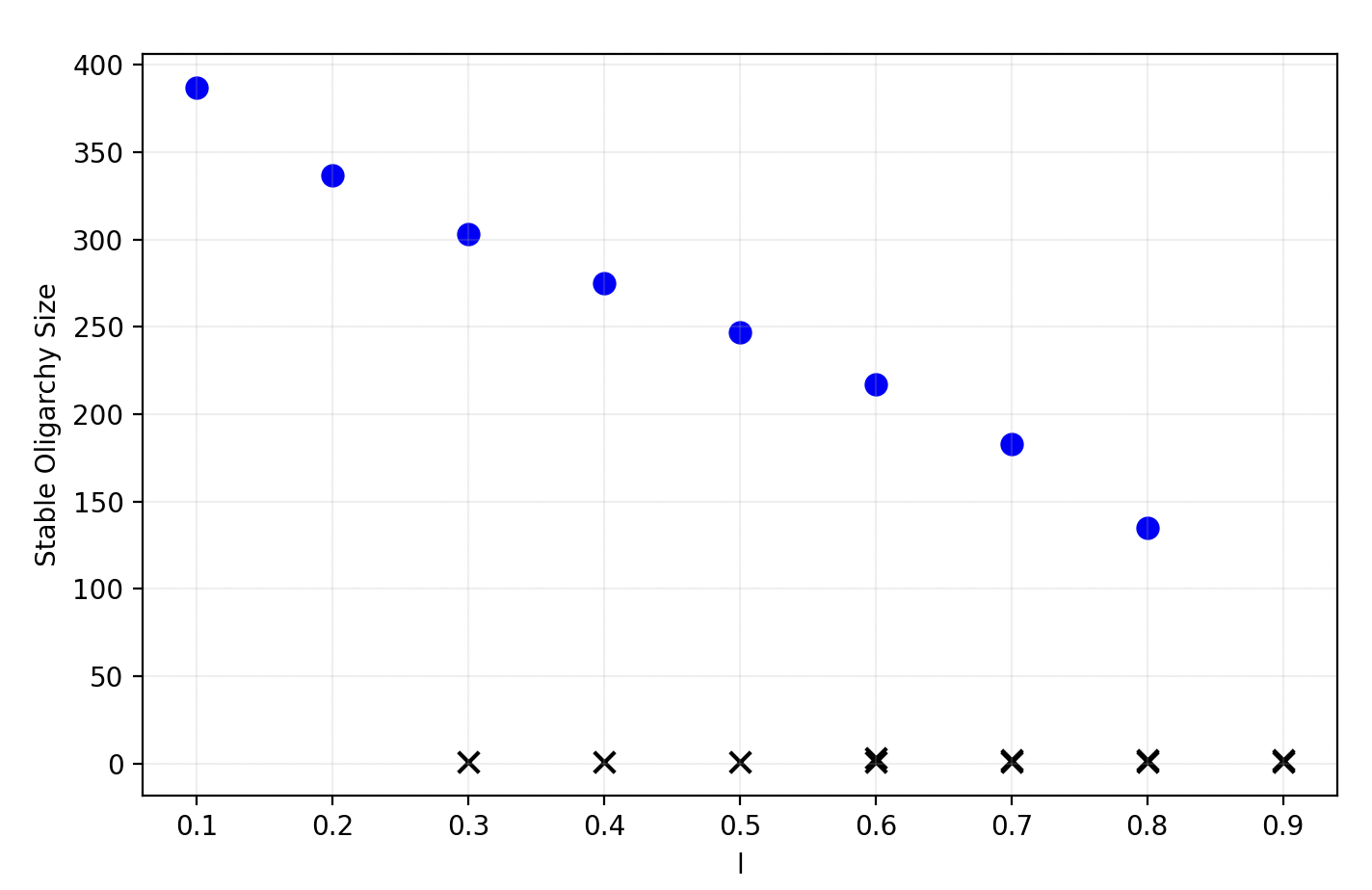}
        \label{fig:slice_p}
    }
    \hfill
    \subfloat[Stable oligarchies as a function of $p$ with $\lambda=0.5$]{
        \centering
        \includegraphics[width=0.48\textwidth]{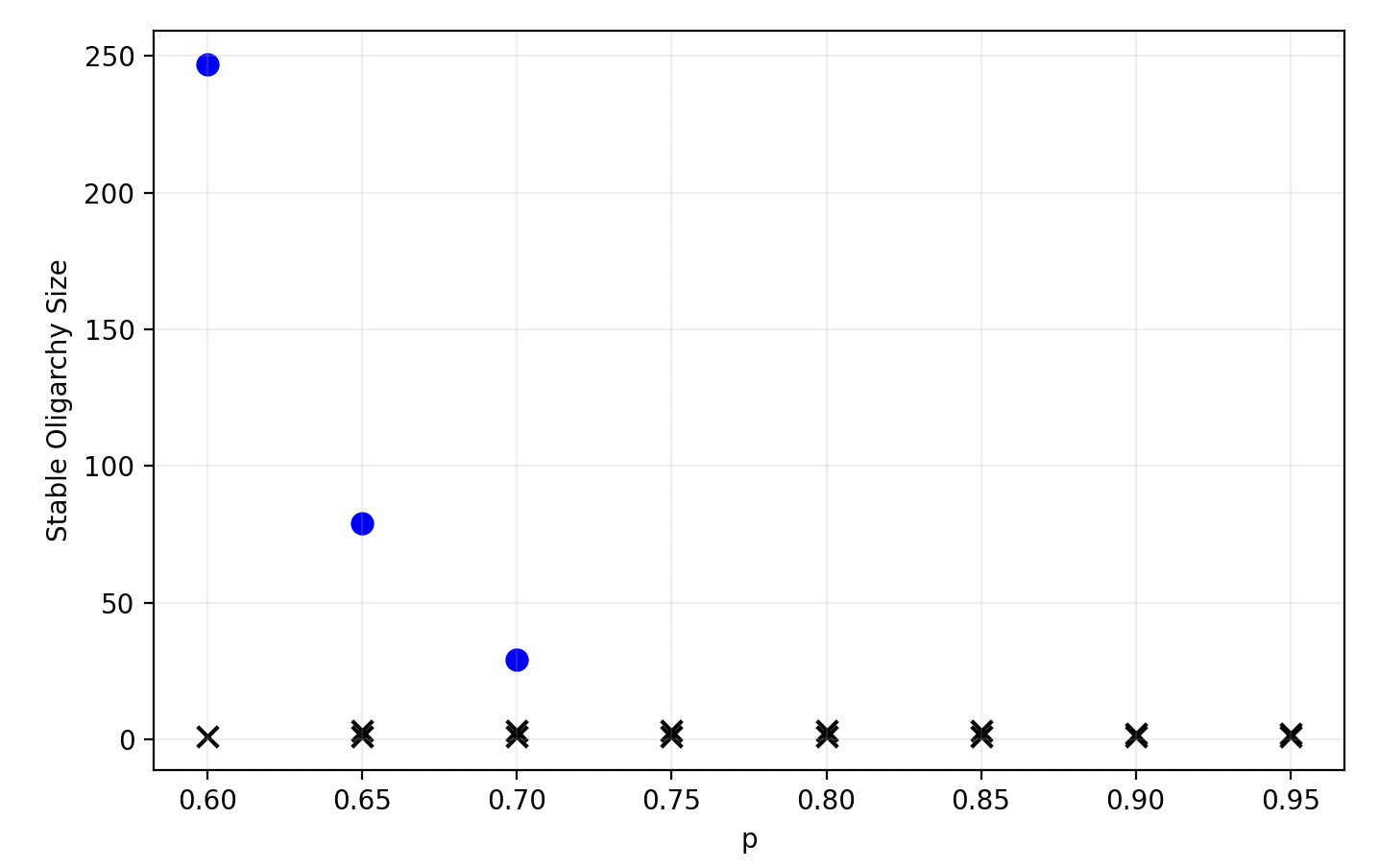}
        \label{fig:slice_l}
    }
        
    \caption{Two-dimensional Slices of the three-dimensional graph of stable-oligarchies as a function of $\lambda$ and $p$. Black $x$ marks are used for small oligarchies (such as dictatorship, $2$-oligarchy and $3$-oligarchy), and blue circles for larger-value oligarchies. }
    \label{fig:slices}
\end{figure}

We include ``slices'' of the graph at both $\lambda = 0.5$, and $p=0.6$ in Figure~\ref{fig:slices}. 

Beyond the self-maintaining points, we are also interested in dynamics. We consider $p=0.6$ with $\lambda \in \{0.6,0.7,0.8,0.9\}$. As in Figure~\ref{fig:slice_p}, there are $3$ equilibria (a dictatorship, a small oligarchy, and a large committee) with $\lambda \in \{0.6,0.7,0.8\}$, and only a dictatorship and a small oligarchy with $\lambda = 0.9$. We wish to understand where do we end up if we start in an arbitrary oligarchy. Figure~\ref{fig:multi_l_dynamics} shows the answer. 

\begin{figure}[htbp] 
    \centering
    \includegraphics[width=0.8\textwidth]{figures/dynamics.png}
    \caption{Centralization dynamics for different extraction Levels ($l \in \{0.6, 0.7, 0.8, 0.9\}$). The colored ribbons represent basins of attraction for the Dictatorship (green), Oligarchy (red), and Committee (blue), with purple indicating that both Committee and Oligarchy can be reached. The black line represents the utility function $u_i(O_i)$, with stable rules marked as gold points.}
    \label{fig:multi_l_dynamics}
\end{figure}

We note the following. As $\lambda$ increases, the tendency for centralization shows up not only in terms of the self-maintaining committee being of smaller $i$ value (and disappearing as a self-maintaining solution with $\lambda = 0.9$), but also in terms of the dynamics: We have more red lines instead of purple (leading only to a small oligarchy rather also possibly to the large committee) and more purple lines instead of blue. The dictatorship is isolated and only leads to itself. We can see why for $\lambda = 0.6$, high values of $i$ lead only to the large committee: Since the function is overall increasing\footnote{The increase has some ``waviness'' depending on whether $i$ is odd or even, due to the way tie-breaking is handled.} after reaching the global minimum and up to the self-maintaining point at $i=217$, then high enough values of $i \leq 217$ are Halving-Safe but not Right-Maximal, and so they converge to the large committee. Values of $i > 217$ are not Halving-Safe (the function is slightly decreasing in that range). 

\section{Practical Takeaways Of Our Results For DAOs}

We draw several practical conclusions from our results to DAOs. 
These takeaways should serve as a rough structural metaphors: DAOs typically have many institutional idiosyncrasies that obscure baseline governance dynamics. Take for example Optimism, governed by the Optimism Collective DAO. Its operating manual specifies a bicameral structure, composed of a "Token house" (for regular proposals) and a "Citizens' House" for governance changes. Proposals are categorized into types, and different voting rules apply to each type. With each proposal, a type-dependent quorum of participation is required, and a different kind of majority. Besides the bicameral structure, there is also a ``Veto'' power which is meant to be reserved for rare circumstances. Voters are divided into different stakeholder groups and some percentage of voters within a group must agree to veto a decision. There are also processes controlling how a proposal is proposed and brought to a vote. 
Our model and extensions allow understanding some baseline mechanics, so to allow mechanism and governance designers to better anticipate the natural trajectory of a DAO and intentionally construct primitives to counteract unwanted centralization.

\begin{itemize}
    \item \textit{Baseline model and forward-looking voters}. Our baseline model shows that self-maintaining rules are heavily centralized, unless beliefs are optimistic. While we do not model an important aspect of DAOs, namely the fact that tokens are tradable and can also be delegated, this is an un-ignorable structural force: Trades may change current stakeholders and power balance, but centralization forces continue. A cautious planner that wishes to avoid a dictatorship may choose to set a more open (but still very centralized) self-maintaining rule, such as a $3$-oligarchy or quasi-dictatorship. If the number of stakeholders is small, and they are considered farsighted,
    our computational results in Section~\ref{sec:forward_looking} suggest that computing forward-looking self-maintaining SCFs in a given system is feasible and fruitful in terms of coming up with new usable SCFs that have good social and Nash welfare relative to the pessimistic self-maintaining SCFs. Our general algorithm can be used for other belief assumptions and domains. 

    \item \textit{Bi-level constitutions.} Our discussion of bi-level constitutions suggests that many natural WMGs can achieve high-level of self-maintenance (pessimistic) if coupled with Unanimity voting to change it. In particular, this includes simple majority voting. Bi-level constitutions are not common in DAOs (a notable exception is Optimism \cite{optimismManual} previously mentioned), but our results show they are a useful primitive. 

    \item \textit{Expansion and Retraction of Voting Rights.} We see that centralization is increasing with how extractive the economy is. For DAOs, this means that if the balance between \textit{current locked-up funds} and \textit{expected future revenue} shifts towards the locked-up funds, we could expect centralization of power. 
Shocks in the decision-making quality of individuals also relates to centralization. For example, if Artificial Intelligence (AI) enhances voters ability to decide, the stable outcome may be a much smaller elite size. 
Of particular interest are our dynamics results, where we 
identify a critical sensitivity to initial conditions: different starting states converge to distinct stable outcomes. This suggests that achieving an inclusive steady-state requires a highly distributive initial token allocation. This can be achieved through ``airdrops'' (free distribution of tokens to prospective holders) or Initial Coin Offerings (ICOs). Existing economic works indeed thinks of these methods as a commitment device by platforms to not exploit users \cite{goldsteinCommitment,sockinCommitment,Reuter2024Platform}. While in our model it is not guaranteed that the platform stays as decentralized as after the airdrop due to the social choice dynamics, it helps with the equilibrium choice of the more decentralized stable large committee. 
\end{itemize}

\section{A Summary of Our Baseline Model Results}
\label{app:tables_summary}

\begin{table}[htbp]
\centering
\caption{Our Results for General Boolean Functions}
\renewcommand{\arraystretch}{1.5}
\begin{tabular}{|p{3cm}|p{5cm}|p{6cm}|}
\hline
& \textbf{Worst-Case Tie-Breaking} & \textbf{SQP Tie-Breaking} \\
\hline
Optimistic Approach & 
Precisely the proper rules \newline (Theorem~\ref{thm:opt_arbitrary}) & 
Precisely proper or null-preserving rules (Theorem~\ref{thm:opt_SQP}) \\
\hline
Pessimistic Approach & 
The constant-0 function \newline (Theorem~\ref{thm:pess_arbitrary}) & 
Dictatorship \newline 
3-oligarchy \newline 
quasi-dictatorship \newline
(Theorem~\ref{thm:pess_SQP_char})
\newline\newline 
A rule must \underline{be}: proper, null-preserving, downward-closed, bounded monotonicity violation, bounded downward sensitivity. \newline
(Theorem~\ref{thm:pess_SQP_necessary})
\newline \newline
\textbf{A gap exists} between the necessary and sufficient conditions. \newline 
(Example~\ref{ex:gap})
\\
\hline
Unbiased I.i.d. & 
Precisely the constant-0 function, dictatorship, and anti-dictatorship. \newline (Theorem~\ref{thm:unbiased_voters_char}) & 
\\
\hline
\end{tabular}
\end{table}

\begin{table}[htbp]
\centering
\caption{Our Results for Simple Games}
\renewcommand{\arraystretch}{1.5}
\begin{tabular}{|p{3cm}|p{5cm}|p{6cm}|}
\hline
& \textbf{Worst-Case Tie-Breaking} & \textbf{SQP (Best-Case) Tie-Breaking} \\
\hline
Optimistic Approach & 
Precisely the proper rules \newline (Theorem~\ref{thm:opt_arbitrary}) & 
All rules (Theorem~\ref{thm:opt_SQP})\\
\hline
Pessimistic Approach & 
None (Theorem~\ref{thm:pess_arbitrary}) & 
Precisely: \newline
Dictatorship \newline 
3-oligarchy \newline 
quasi-dictatorship \newline 
(Theorem~\ref{thm:pess_SQP_char})
 \\
\hline
Unbiased I.i.d. & 
Precisely dictatorships \newline (Theorem~\ref{thm:unbiased_voters_char})& 
 \\
\hline
\end{tabular}
\end{table}

\begin{table}[htbp]
\centering
\caption{Our Results for Weighted Majority Games}
\renewcommand{\arraystretch}{1.5}
\begin{tabular}{|p{3cm}|p{5cm}|p{6cm}|}
\hline
& \textbf{Worst-Case Tie-Breaking} & \textbf{SQP (Best-Case) Tie-Breaking} \\
\hline
Optimistic Approach & 
All rules (Theorem~\ref{thm:opt_arbitrary}) & 
All rules (Theorem~\ref{thm:opt_SQP}) \\
\hline
Pessimistic Approach & 
None (Theorem~\ref{thm:pess_arbitrary})& 
Precisely: \newline
Dictatorship \newline 
3-oligarchy \newline 
quasi-dictatorship \newline
(Theorem~\ref{thm:pess_SQP_char})
 \\
\hline
Unbiased I.i.d. & 
Dictatorships (Theorem~\ref{thm:dictatorship}) \newline  Maximal quasi-dictatorship \newline (Example~\ref{ex:maximal_quasi_dictatorship})& 
 \\
\hline
\end{tabular}
\end{table}

\begin{figure}
    \centering
    \includegraphics[width=\linewidth]{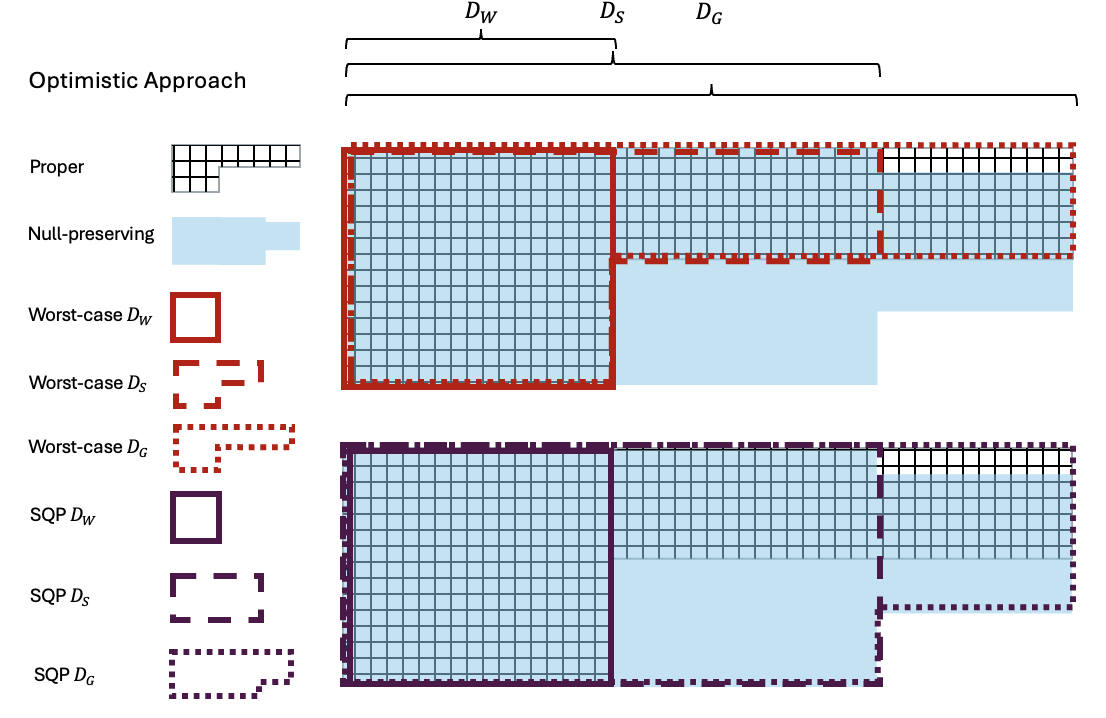}
    \caption{Venn diagram representation of our optimistic approach results. For readability, the results are separated to these using worst-case tie-breaking (the upper diagram) and SQP tie-breaking (the lower diagram).  }
    \label{fig:optimistic_venn}
\end{figure}

\begin{figure}
    \centering
    \includegraphics[width=\linewidth]{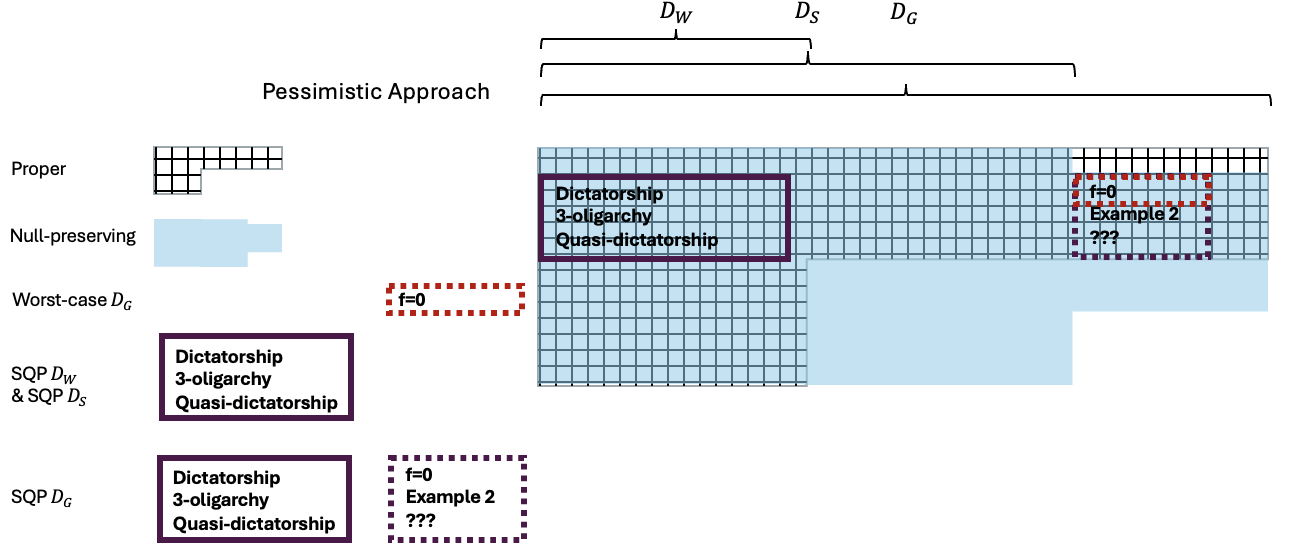}
    \caption{Venn diagram representation of our pessimistic approach results. The worst-case tie-breaking characterization for $D_S$ and $D_W$ are the empty sets, and so are omitted from the diagram. ``???'' is included where the characterization is not tight. }
    \label{fig:pessimistic_venn}
\end{figure}

\begin{figure}
    \centering
    \includegraphics[width=\linewidth]{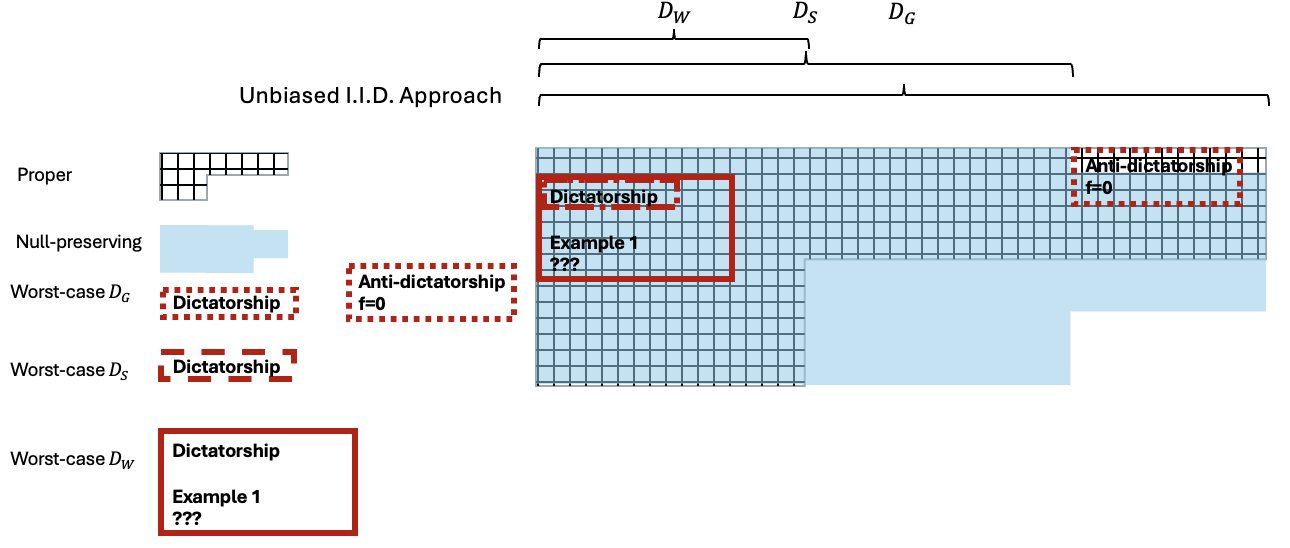}
    \caption{Venn diagram representation of our unbiased i.i.d. approach results. ``???'' is included where the characterization is not tight. All SQP tie-breaking characterizations are omitted from the diagram: We know only that they are weakly more permissive than the corresponding worst-case tie-breaking characterizations. 
     }
    \label{fig:unbiased_iid_venn}
\end{figure}

\newpage

\section{Missing Proofs for the Optimistic and Pessimistic Cases}

\OptArbitrary*

\begin{proof}

(Characterization $\implies$ self-maintaining)

Consider some other SCF $f'$. Let $i^*$ be the first index $i$ in the ordering that is used by the lexicographic distribution $F$ so that $f'(v^i) \neq f(v^i)$. For every voter $j$ so that $v^{i^*}_j = f(v^{i^*})$, we have $u_j(f, F) > u_j(f', F)$, since $F$ is lexicographic. For the same reason, for every voter $j$ so that $v^{i^*}_j = f'(v^{i^*})$, we have the opposite: $u_j(f', F) > u_j(f, F)$. This yields $\mathbf{c} \in \{v^{i^*}, \neg v^{i^*}\}$. More specifically, if $f(v^{i^*}) = 0$, it is $v^{i^*}$, and if $f(v^{i^*}) = 1$, it is $\neg v^{i^*}$). In the former case, we conclude that $f(\mathbf{c}(f, f')) = f(v^{i^*}) = 0$, and $f$ is self-maintaining. In the latter case, by our theorem's assumption that $f(v) = 1 \implies f(\neg v) = 0$, and since $f(v^{i^*}) = 1$, we conclude that $f(\mathbf{c}(f, f')) = f(\neg v^{i^*}) = 0$, and $f$ is self-maintaining. 

(Does not follow characterization $\implies$ not self-maintaining)

We first remark that all weighted majority games in $D_W$ are proper, so no $f$ would satisfy this condition, and so this implication holds in a vacuous sense. 

Consider $f$ so that there is some $v$ with $f(v) = f(\neg v) = 1$. If $f$ is monotone, w.l.o.g., we choose such $v$ that is a minimal winning coalition (we can start with some arbitrary $v$ that satisfies the property, take a voting preference $v'$ so that $S(v')$ is a minimal winning coalition, and for $\neg v'$, $S(\neg v) = N \setminus S(v) \subseteq N \setminus S(v') = S(\neg v')$, and so by monotonicity $f(\neg v') \geq f(\neg v) = 1$). If $f$ is unanimity-preserving, then $v \neq \mathbf{0}$, as $f(\mathbf{0}) = 0$, and $v \neq \mathbf{1}$, as $f(\neg \mathbf{1}) = 0$. 

Let $F$ be any distribution over $\Omega$. 

Consider $f'$ so that $f'(v) \neq f(v)$, and for all other $v'$, $f'(v') = f(v')$. If $f$ is monotone, and we choose $v$ so that $S(v)$ is a minimal winning coalition, then $f'$ is also monotone. Moreover, since $v \not \in \{\mathbf{0}, \mathbf{1}\}$, our change is still unanimity-preserving. 

We have that $u_j(f') \geq u_j(f)$ whenever $v_j = 0 \neq 1 = f(v)$, and $u_j(f') \leq u_j(f)$ whenever $v_j = 1 = f(v)$. By worst-case tie-breaking we can thus choose $\mathbf{c}(f, f') = \neg v$, and so 
$f(\mathbf{c}(f, f')) = f(\neg v) = 1$, and $f$ is not self-maintaining for $F$.

\qed \end{proof}

\OptSQP*

\begin{proof}

(Characterization $\implies$ self-maintaining)

As noted above, Theorem~\ref{thm:opt_arbitrary} implies that any proper SCF is optimistically self-maintaining with worst-case tie-breaking, and thus also with SQP tie-breaking. We complete the case of an SCF $f$ that is null-preserving. Consider a distribution $F = \mathbf{0} w.p. 1$. For any $f'$, we have $u_i(f, F) \geq u_i(f', F)$ for all voters $i$ (due to $f$ being null-preserving), and so $\mathbf{c}(f,f') = \mathbf{0}$, and by $f$ being null-preserving, we are guaranteed that $f(\mathbf{c}(f,f')) = f(\mathbf{0}) = 0$. 

(Self-maintaining $\implies$ Characterization)

We first remark that all simple voting games (and in particular, all weighted majority games) are null-preserving, so the implication holds in a vacuous sense, and we do not need to care for our construction to be domain-preserving.

Consider $f$ which is self-maintaining for some distribution $F$. If $f$ is proper, we are done, and so we assume it is not, and there is such $\mathbf{v}$ so that $f(\mathbf{v}) = f(\neg \mathbf{v}) = 1$. We also assume, towards contradiction, that $f(\mathbf{0}) = 1$. 

Consider $f'(\mathbf{v}) = 0$ and otherwise $f' = f$. For a distribution $F$ that has $\mathbf{v} \not \in supp F$, $u_i(f, F) = u_i(f', F)$ for all voters $i$, and so $f(\mathbf{c}(f, f')) = f(\mathbf{0}) = 1$. For a distribution $F$ that has $\mathbf{v} \in supp F$, $u_i(f, F) > u_i(f', F)$ for all voters with $v_i = 1$, and $u_i(f, F) < u_i(f', F)$ for all voters with $v_i = 0$, and so $\mathbf{c}(f, f') = \neg \mathbf{v}$. Thus, $f(\mathbf{c}(f, f')) = f(\neg \mathbf{v}) = 1$. 
\qed \end{proof}

\PessArbitrary*

\begin{proof}
Clearly, $f = 0$ is self-maintaining as regardless of voting profile $\mathbf{c}$, the alternative SCF $f'$, the distribution $F$, and the tie-breaking choice induce, we have $f(\mathbf{c}) = 0$. 

Otherwise, it must be that there is some $v^*\in \Omega$ with $f(v^*) = 1$. For the domains $D_S, D_W$ that are \textit{unanimity-preserving}, we can further require that $v^* = \mathbf{1}$. 

Consider $F$ so that $Pr_{\hat{v}\sim F}[\hat{v} = v^*] = 1$. Consider some $v \neq v^*$. In particular, for a function in $D_S$, we can assume $v$ is such that $S(v)$ is a minimal winning coalition, or (in the specific case where $f$ is Unanimity), we assume $S(v)$ is a maximal losing coalition, since we want $v \neq v^*$. In either case, we are guaranteed that $v \neq \mathbf{0}$. 
Then, consider $f'$ so that $f'(v) \neq f(v)$, and for all other vectors $v'$, $f'(v') = f(v')$. By our choice of $v$ for $D_S$, this preserves monotonicity (since we change the outcome of a voting preference that corresponds either to a voting minimal winning coalition or a maximal losing coalition), and also unanimity-preserving (since $v \not \in  \{ \mathbf{0}, \mathbf{1} \}$). 

Our proof so far treated $D_S$ and $D_G$, but not $D_W$. 
For a function $f$ in $D_W$, let $\mu$ be its minimal quora, i.e., the size of its smallest minimal winning coalition. If $\mu = n$, consider $\mu' = \lfloor \frac{n}{2} \rfloor + 1$, otherwise, consider $\mu' = n$. Let $w'_1 = \ldots = w'_n = 10$, $T = \mu' \cdot 10 - 1$, and let $f'$ be the corresponding Boolean function of the weighted majority game. Notice that $\mu'$ is the minimal quora of $f'$, and so it must be that $f \neq f'$. 

Finally, for all our domains $D_W, D_S, D_G$, we gather that by our different constructions it holds that $f(v^*) = f'(v^*)$, it holds under $F$ that for all voters $i$,
$$u_i(f') = u_i(f)$$

(as all other voting preferences are not in the support of $F$). Thus, by worst-case tie-breaking, we can choose $\mathbf{c}(f, f') = v^*$, and have
$f(\mathbf{c}(f, f')) = f(v^*) = 1$. 

\qed \end{proof}

\begin{definition}
For a voting profile $v_i$, we let $S_i = \{ j | v^i_j = 1\}$ be the set of voters that support the decision under $v_i$.

    \textit{Downward closed.} For any $v^1, v^2 \in \Omega$ with $v^1 \neq v^2$, consider $v$ so that $v_j = 1$ if and only if $j \in S_1 \cap S_2$. Then, if $f(v^1) = f(v^2) = 0$, then so is $f(v) = 0$. In words, if two voting profiles are ruled as \textit{against} under the SCF $f$, then the voting profile where the voters supporting the motion are the intersection of supporters on both other profiles is also ruled against. 

    \textit{Bounded Monotonicity Violation.} For any two voting profiles $v^1, v^2 \in \Omega$ so that $S_1 \subset S_2$, and $f(v^1) = 1, f(v^2) = 0$, then for any $v^3$ so that $(S_2 \setminus S_1) \cap S_3 = \emptyset, S_1 \subset S_3$, we have $f(v^3) = 1$. In words, if the SCF violates monotonicity for $v^1, v^2$, in the sense that $v^1$ has a strict subset of supporting voters but is accepted while $v^2$ is not, then any other $v^3$ that extends $v^1$ (in terms of supporting voters) is accepted. 

    \textit{Bounded Downward Sensitivity.}
For any $v^1$ with $|S_1| \geq 3$ and $f(v^1) = 1$, there are at most $2$ vectors $v^2, v^3$ with $S_2 \subset S_1, S_3 \subset S_1$ and $|S_2| = |S_3| = |S_1| - 1$ so that $f(v^2) = f(v^3) = 0$. In words, if a voting profile with $k$ voters in favor of the motion passes under $f$, then removing one voter should also pass, with the exception of possibly two voters that when removed fail the vote.  

\end{definition}

Notice that monotone SCFs are naturally downward closed and have bounded monotonicity violation. Thus, the following theorem's main significance is for the general Boolean function domain (where it shows the SCF must be ``almost'' monotone), and the bounded downward sensitivity property, which is relevant to all domains. 


\begin{restatable}[Necessary Conditions for Pessimistic Self-Maintenance with SQP tie-breaking]{theorem}{PessSQPNecessary}

\label{thm:pess_SQP_necessary}

Any SCF $f$, that is pessimistic self-maintaining with SQP tie-breaking for any of the domains $D_G, D_S, D_W$, is proper, null-preserving, downward-closed, has bounded monotonicity violation and bounded downward sensitivity. 
\end{restatable}

\begin{proof}

We first note that since monotonicity implies downward-closed, all SCFs in $D_S, D_W$ satisfy it, so it holds as a necessary condition in a vacuous sense. Similarly, null-preserving and proper hold immediately for $D_S, D_W$ (and thus for these properties, we focus on the $D_G$ case). 

(Proper)

Notice that this is the necessary (and sufficient) condition in Theorem~\ref{thm:opt_arbitrary}. However, we now consider on one hand a pessimistic (rather than optimistic) approach, which is stricter, but, on the other hand, we use SQP tie-breaking rather than arbitrary tie-breaking, which is looser. 

If there is some $v$ so that $f(v) = f(\neg v) = 1$, then consider $F$ so that $Pr_{\hat{v}\sim F}[\hat{v} = v] =1$. Let $f'$ so that $f'(v) = 0 \neq 1 = f(v)$, and for any other $v'$, $f'(v') = f(v')$. Then for any voter $i$ with $v_i = 0$, we have $u_i(f) = 0 < 1 = u_i(f')$, and for any voter $i$ with $v_i = 1$, we have $u_i(f) = 1 > 0 = u_i(f')$, and so $\mathbf{c}(f,f') = \neg v$. Then, $f(\mathbf{c}(f,f')) = f(\neg v) = 1$, in contradiction to self-maintenance.

(Null-preserving)

Consider $F$ so that for some $v\in \Omega$, $Pr_{\hat{v}\sim F}[\hat{v} = v] = 1$, and consider $f'$ so that for some $\tilde{v} \neg v$, $f'(\tilde{v}) = 1 - f(\tilde{v})$, and for any other $v'$, $f'(v') = f(v')$ (in particular, $f'(v) = f(v)$). Then, for any voter $i$, $u_i(f) = u_i(f')$, and by SQP tie-breaking, $\mathbf{c}(f, f') = \mathbf{0}$. We then have by self-maintenance that $f(\mathbf{0}) = 0$. 

(Downward-closed) We first note that since monotonicity implies downward-closed, all SCFs in $D_S, D_W$ satisfy it, so it holds as a necessary condition in a vacuous sense. 

Consider some $v^1, v^2, v$ and their respective sets of indices where the voter votes $1$, $S_1, S_2, S_1 \cap S_2$. We have that $f(v^1) = f(v^2) = 0$. Let $F$ so that $Pr_{\hat{v}\sim F}[\hat{v} = v^1] = Pr_{\hat{v}\sim F}[\hat{v} = v^2] = \frac{1}{2}$. Let $f'$ be an SCF with $f'(v^1) = f'(v^2) = 1$, and for any other $v'$, $f'(v') = f(v')$. Then, for any voter $i$ in $S_1 \cap S_2$, the voter has $v^1_i = v^2_i = 1$, and thus $u_i(f) = 0 < 1 = u_i(f')$, and thus $c_i(f, f') = 1$. For any voter $i$ in $S_1 \setminus S_2$ (or, symmetrically, in $S_2 \setminus S_1$), we have $v^1_i = 1, v^2_i = 0$, and so $u_i(f) = \frac{1}{2} = u_i(f')$, and by SQP tie-breaking, $c_i(f, f') = 0$. For any voter $i$ in $[n] \setminus (S_1 \cup S_2)$, we have $v^1_i = v^2_i = 0$, and so $u_i(f) = 1 > 0 = u_i(f')$, and $c_i(f, f') = 0$. Overall, $f(v) = f(\mathbf{c}(f, f')) = 0$ by the self-maintenance condition.  

(Bounded monotonicity violation) We first note that since monotonicity implies bounded monotonicity violation, all SCFs in $D_S, D_W$ satisfy it, so it holds as a necessary condition in a vacuous sense.

We have $S^3 \cap S^2 = S^1$, so by downward-closed, if $f(v^3) = f(v^2) = 0$, then so must be $f(v^1) = 0$, in contradiction. This is essentially a different way to state downward-closed, that emphasizes the implications of too many instances of non-monotonicity of the outcome in the number of voters voting $1$. 

(Bounded downward sensitivity) 

Let some $v^1$ with $|S_1| \geq 3$, $f(v^1) = 1$, and assume towards contradiction there are $3$ vectors $v^2,v^3,v^4$ with $|S_2| = |S_3| = |S_4| = |S_1| - 1 \geq 2$ and $f(v^2) = f(v^3) = f(v^4) = 0$. Consider $F$ so that $Pr_{\hat{v}\sim F}[\hat{v} = v^2] = Pr_{\hat{v}\sim F}[\hat{v} = v^3] = Pr_{\hat{v}\sim F}[\hat{v} = v^4] = \frac{1}{3}$.

If $f \in D_G \setminus D_S$,
consider $f'$ so that $f'(v^2) = f'(v^3) = f'(v^4) = 1$, and for any other $v'$, $f'(v') = f(v')$. 

If $f\in D_S$, construct the following WMG: et $w_i = 2$ for any $i\in S_1$, and for any other $i$, $w_i = 0$. Let $T = 2 \cdot |S_1| - 3$, and let $f'$ be the corresponding Boolean function. Notice that since $|S_1| \geq 3$, $T$ equals at least half of the sum of weights. Since $f' \in D_W$, it is domain-preserving for both $D_S, D_W$. 

For any voter $i \in S_1$, $u_i(f) \leq \frac{1}{3}$ (it is $\frac{1}{3}$ if for some $k \in \{2,3,4\}$, $S_1 \setminus S_k = \{i\}$, and $0$ otherwise), and $u_i(f') \geq \frac{2}{3}$. For any other voter $i$, $u_i(f) = 1 > 0 = u_i(f')$. Thus, $f(\mathbf{c}(f,f')) = f(v^1) = 1$, in contradiction to self-maintenance.

\qed \end{proof}

\begin{figure}[htbp]
    \centering
    \begin{minipage}{0.48\textwidth}\includegraphics[width=\linewidth]{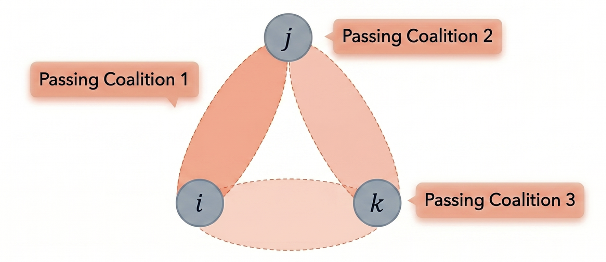}
    \caption{The minimal winning coalitions of a $3$-oligarchy: Every pair of the voters $i,j,k$ that constitute the oligarchy.}
\label{fig:olig3}
    \end{minipage}\hfill
\begin{minipage}{0.48\textwidth}
\includegraphics[width=\linewidth]{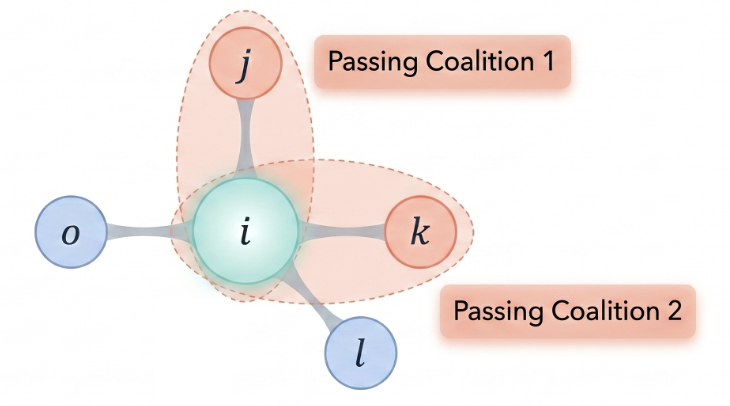}
    \caption{The schematics of the minimal winning coalitions of a quasi-dictatorship: All the minimal winning coalitions are pairs that contain the voter $i$.}
\label{fig:pdp}
        \end{minipage}
\end{figure}

\MinimalWinningTwo*

\begin{proof}
Consider a proper monotone SCF $f$ where all the minimal winning coalitions are of size $2$ and is not a $3$-oligarchy or a quasi-dictatorship. Note that in a proper monotone SCF, no two minimal winning coalitions can be disjoint. 

Since $f$ is not a quasi-dictatorship, it means that for voter $i$ that belongs in a minimal winning coalition, there is a minimal winning coalition $(i',j')$ that $i$ does not belong to. 

We claim that it must be that there is a minimal winning coalition $(i,j)$ so that both voters are different from another minimal winning coalition $(i',j')$. 

Assume otherwise. There is some minimal winning coalition that contains $i$. If the pairs $(i,i')$ and $(i,j')$ are minimal winning coalitions, then this is a $3$-oligarchy. If $i$ is part of two minimal winning coalitions but not the above, then at least one of the pairs has two voters different than $(i',j')$ (namely, $i$ and the voter it is paired with). Otherwise, 
(w.l.o.g.) $(i,i')$ is the unique minimal winning coalition that contains $i$. Since the SCF is not a quasi-dictatorship, there is some pair that does not contain $i'$. Then for this minimal winning coalition,  both voters are not $i, i'$ (since $(i,i')$ is the only pair where $i$ appears). 

This yields two disjoint minimal winning coalitions of $f$, in contradiction. 
\qed \end{proof}

\PessSQPChar*

\begin{proof}

(The specified rules are self-maintaining)

For a dictatorship $f$ of voter $i$, it has $u_i(f, F) = 1$ for any distribution $F$. Thus, by SQP tie-breaking, $c_i(f, f') = 0$ for any $f'$, and so $f(\mathbf{c}(f,f')) = c_i(f,f') = 0$.

We next consider a $3$-oligarchy. Let $S$ be the oligarchy voters. 

The key property is that for any $F$ and $i,j \in S$, $u_i(f, F) + u_j(f, F) = \max_{f'} u_i(f', F) + u_j(f', F)$. This holds because the outcome follows $i$ and $j$ consensus whenever such exists. 
Thus, for any $f'$ and $i\in S$ so that $u_i(f', F) > u_i(f, F)$, it must be that for any $j \in S, j\neq i$, $u_j(f, F) > u_j(f', F)$. Thus, the choice vector $\mathbf{c}(f, f')$ will always have a majority $0$ among the voters in $S$, and so we must have $f(\mathbf{c}(f, f')) = 0$, and so self-maintenance is satisfied. 

The argument to show that a quasi-dictatorship $f$ is self-maintaining is very similar: If we let $i$ be the unique voter that appears in all pairs $S(f)$ (the ``quasi-dictator''), then the only vectors $v$ so that $f(v) = 1$ are ones where $v_i = 1$ and $v_j = 1$ for some $(i,j) \in S(f)$, and the key property holds for $i,j$. Since this holds for any $j$ that is in a minimal winning coalition with $i$, 
for any $F$ to have $u_i(f', F) > u_i(f, F)$ (in order for $c_i(f, f') = 1$), it must imply by the key property $u_j(f', F) < u_j(f, F)$ for any paired voter $j$, which then implies $f(\mathbf{c}(f,f')) = 0$. 

(No other rules are self-maintaining in $D_S, D_W$)

By Theorem~\ref{thm:pess_SQP_necessary}, being proper is a necessary condition for self-maintenance. All rules in $D_S$ and $D_W$ are monotone. Thus, by Lemma~\ref{lem:3olig_pdp}, any rule where all the minimal coalitions are of size at most $2$ is one of the specified rules. Now, assume towards contradiction a rule with a minimal winning coalition $W$ of size at least $3$. By Theorem~\ref{thm:pess_SQP_necessary}, a necessary condition is bounded downward sensitivity, which means a strict subset of $W$ is winning, in contradiction. 

\qed \end{proof}

The following example shows a gap exists between the necessary and sufficient conditions for the domain $D_G$: 

\begin{example}
\label{ex:gap}
For any $n\geq 3$, consider the following SCF:
$$f(v) = \begin{cases} 1 & (v_1 = 1 \land v_2 = 0) \\
0 & Otherwise.\end{cases}$$

It is clear that $f$ does not fall within our class of sufficient characterization, since $f(\mathbf{1}) = 0$. However, it is self-maintaining, since if $f(v) \neq v_1$, it must hold that $v_1 = v_2$. Thus, for any $f' \neq f$, it holds that $u_1(f') - u_1(f) \leq u_2(f') - u_2(f)$ (changing any $v$ where voters $1$ and $2$ agree equally changes their utility. For all other $v$, it holds that $f(v) = v_1$, so any change decreases voter $1$'s utility, and may equally decrease voter $2$'s utility or leave it the same). 

This means that to have $c_1(f,f') = 1$, we must have $u_1(f') - u_1(f) > 0$, and so we must have $u_2(f') - u_2(f) > 0$ and $c_2(f,f') = 1$. However, for any vector with $v_1 = v_2 = 1$ we have $f(v) = 0$. 

\end{example}

\section{Missing Proofs for The Unbiased I.I.D. Case (Section~\ref{sec:iid})}

Consider i.i.d. preference voters with parameter $p$. First, we can identify the following simple self-maintaining rules:

\begin{restatable}{theorem}{Dictatorship}
\label{thm:dictatorship}
With i.i.d. preference voters and worst-case tie-breaking, any $i$-dictatorship, $i$-anti-dictatorship, and the constant $0$ are self-maintaining. 
\end{restatable}

\begin{proof}

The constant SCF $f$ is self-maintaining, as regardless of the voters' preferences of replacing it with another SCF $f'$, it outputs $0$, satisfying the self-maintenance constraint.  

We show that $i$-dictatorship is self-maintaining. 

We can directly calculate: $$u_i(f^i_{dict}) = Pr_{v_j \sim F_p}[f^i_{dict}(v) = v_i] = 1.$$

Since for any $f$, $1$ is a trivial upper bound over $u_i(f)$ (as it is a probability of some event), 
and, moreover, $i$-dictatorship is the unique SCF that has $u_i(f) = 1$, 
we have $c_i(f^i_{dict}, f') = 0$, for any other $f'$. Then, regardless of $\mathbf{c}_{-i}(f^i_{dict},f')$ (since the $i$-dictatorship output only depends on $v_i$):

$$f^i_{dict}(\mathbf{c}(f^i_{dict},f')) = 0.$$  

The proof that $i$-anti-dictatorship is self-maintaining is similar: Since it is the unique SCF that has $u_i(f) = 0$, we have $c_i(f^i_{anti-dict}, f') = 1$, and thus 
$$f^i_{anti-dict}(\mathbf{c}(f^i_{anti-dict},f')) = 0.$$ 
\qed \end{proof}

\begin{definition}
\label{def:IIDpartition}
For the general Boolean function case, it is useful to partition the voters into three subsets, depending on their utility level under $f$: 

$$S_1 = \{i | u_i(f) < 2p(1-p)\}, S_2 = \{i | 2p(1-p) \leq u_i(f) < p^2+(1-p)^2\}, S_3 = \{i |  u_i(f) \geq p^2+(1-p)^2\}.$$

We use the notation $S_1, S_2, S_3$ in the lemmas and theorems that follow. 

\end{definition}

We characterize a necessary condition for any SCF other than the ones in Theorem~\ref{thm:dictatorship}. 

\begin{restatable}[]{lemma}{MainStructuralLemma}
\label{lem:main_structural}
With unbiased i.i.d. preference voters, any self-maintaining SCF $f$ that is not a dictatorship, anti-dictatorship, or the constant $0$, 
must satisfy the following property: If $S_1 \neq \emptyset, S_3 \neq \emptyset$, then for every vector $v$ that has $v_i = 1$ for any $i\in S_1 \cup S_2$, $f(v) = 0$. For any vector $v$ that has $v_i = 0$ for any $i\in S_2 \cup S_3$, $f(v) = 0$. 
\end{restatable}

\begin{proof}

First, notice that for any $0\leq p\leq 1$, $2p(1-p) \leq p^2+(1-p)^2$, as this inequality is a rearrangement of $0 \leq (1-2p)^2$. Thus, the three sets constitute a full partition of the voters. 

Consider $v$ so that $\forall i\in S_1 \cup S_2, v_i = 1$, and $\forall i\in S_3, v_i = 0$. Then, it must be that $f(v) = 0$. Otherwise, if $f(v) = 1$, consider $f'$ to be the $i^*$-dictatorship for some voter in $S_1 \cup S_2$. It has $u_{i^*}(f') = 1$, and $\forall i \in S_1 \cup S_2 \setminus \{i^*\}, u_i(f') = Pr_{v\sim F}[v_i = f(v)] = Pr_{v_i \sim F_p, v_{i^*} \sim F_p}[v_i = v_{i^*}] = \sum_{x=0}^1 Pr_{v_{i^*} \sim F_p}[v_i = v_{i^*} | v_i = x] \cdot Pr_{v_i \sim F_p}[v_i = x] = p^2 + (1-p)^2$. Therefore, $\mathbf{c}_i(f, f') = 1$ for any $i\in S_1 \cup S_2$, as they have $u_i(f) < u_{i}(f')$, and for any $i\in S_3$,  $\mathbf{c}_i(f, f') = 0$, as $u_i(f) \geq u_i(f')$. We then have 
$f(\mathbf{c}(f,f')) = f(v) = 1$, violating self-maintenance. 

Consider $v$ so that $\forall i\in S_1 \cup S_2, v_i = 1$, for some $i^*\in S_3, v_{i^*} = 1$, and for all other $i\neq i^*$ so that $i\in S_3$, $v_i = 0$. Then, it must be that $f(v) = 0$. 
Otherwise, if $f(v) = 1$, consider $f'$ to be the $i^*$-dictatorship. It has $u_{i^*}(f') = 1$, and $\forall i\neq i^*, u_i(f') = p^2 + (1-p)^2$. Therefore, $\mathbf{c}_i(f, f') = 1$ for any $i\in S_1 \cup S_2$, $\mathbf{c}_{i^*}(f,f') = 1$, and $\mathbf{c}_i(f, f') = 0$ for any $i\in S_3 \setminus \{i^*\}$. I.e., $f(\mathbf{c}(f,f')) = f(v) = 1$, violating self-maintenance. 

We now show by induction that for any $v$ so that $\forall i\in S_1 \cup S_2, v_i = 1$, $f(v) = 0$. The induction is on the number $k$ of voters $i\in S_3$ with $v_i = 1$. The $k = 0,1$ base cases were directly shown before. Now consider that $f(v) = 0$ for all vectors $v$ with $k' \leq k$ for some $k \geq 1$, and we wish to show that this holds for any vector $v$ with $k' = k+1$. Consider $v$ so that $v_i = 1$ for all $i\in S_1 \cup S_2$, and there is $S' \subseteq S_3$ of size $k+1$ so that $\forall i\in S', v_i = 1$, and $\forall i\in S_3 \setminus S', v_i = 0$. Let $i_1, i_2 \in S'$ be two different voters in $S'$. Define $v'$ so that $v'_i = 1$ for all $i\in S_1 \cup S_2 \cup (S' \setminus i_1)$, and $v'_i = 0$ otherwise, and $v''$ so that $v''_i = 1$ for all $i\in S_1 \cup S_2 \cup (S' \setminus i_2)$, and $v''_i = 0$ otherwise. By the induction assumption, $f(v') = f(v'') = 0$. Assume towards contradiction that $f(v) = 1$. Consider $f'$ which is identical to $f$, besides $f'(v') = f'(v'') = 1$. Then, the utility of all voters in $S_1 \cup S_2 \cup (S' \setminus \{i_1, i_2\})$ improves (as they have $v'_i = v''_i = 1$, and $f'(v') = f'(v'') = 1$). The utility of voters $i_1, i_2$ does not change:

\[
\begin{split}
& u_{i_1}(f') \\
& = u_{i_1}(f) - Pr_{v\sim F}[v = v'] + Pr_{v\sim F}[v = v''] \\
& = u_{i_1}(f) - p^{|S_1| + |S_2| + k}(1-p)^{|S_3| - k} + p^{|S_1| + |S_2| + k}(1-p)^{|S_3| - k} \\
& = u_{i_1}(f),
\end{split}
\]
and similarly for $u_{i_2}(f')$. The utility of voters $S_3 \setminus S'$ declines (as they have $v'_i = v''_i = 0$, and $f'(v') = f'(v'') = 1$). Overall, by worst-case tie-breaking, we can choose $\mathbf{c}(f, f') = v$, and have $f(\mathbf{c}(f, f')) = f(v) = 1$, in contradiction to self-maintenance. We thus conclude that $f(v) = 0$. 

To prove the second part of the lemma statement, we apply a very similar argument, now using \textit{anti-dictatorship} as the base for our induction. 

Consider $v$ so that $\forall i\in S_2 \cup S_3, v_i = 0$, and $\forall i\in S_1, v_i = 1$. Then, it must be that $f(v) = 0$. Otherwise, if $f(v) = 1$, consider $f'$ to be the $i^*$-anti-dictatorship for some voter in $S_2 \cup S_3$. It has $u_{i^*}(f') = 0$, and $\forall i \in S_2 \cup S_3 \setminus \{i^*\}, u_i(f') = Pr_{v\sim F}[v_i = f(v)] = Pr_{v_i \sim F_p, v_{i^*} \sim F_p}[v_i \neq v_{i^*}] = \sum_{x=0}^1 Pr_{v_{i^*} \sim F_p}[v_i \neq v_{i^*} | v_i = x] \cdot Pr_{v_i \sim F_p}[v_i = x] = 2p(1-p)$. Therefore, $\mathbf{c}_i(f, f') = 1$ for any $i\in S_1$, as they have $u_i(f) < u_{i}(f')$, and for any $i\in S_2 \cup S_3$, 
$\mathbf{c}_i(f, f') = 0$, as $u_i(f) \geq u_i(f')$. We then have 
$f(\mathbf{c}(f,f')) = f(v) = 1$, violating self-maintenance. 

Consider $v$ so that $\forall i\in S_2 \cup S_3, v_i = 0$, for some $i^* \in S_1$, $v_{i^*} = 0$, and $\forall i\in S_1 \setminus \{i^*\}, v_i = 1$. Then, it must be that $f(v) = 0$. Otherwise, if $f(v) = 1$, consider $f'$ to be the $i^*$-anti-dictatorship. It has $u_{i^*}(f') = 0$, and $\forall i \in S_2 \cup S_3 \setminus, u_i(f') = 2p(1-p)$. Therefore, $\mathbf{c}_i(f, f') = 1$ for any $i\in S_1 \setminus \{i^*\}$, as they have $u_i(f) < u_{i}(f')$, and for any $i\in S_2 \cup S_3 \cup \{i^*\}$, 
$\mathbf{c}_i(f, f') = 0$, as $u_i(f) \geq u_i(f')$. We then have 
$f(\mathbf{c}(f,f')) = f(v) = 1$, violating self-maintenance. 

We now show by induction that \textit{every} vector $v$ so that $\forall i\in S_2 \cup S_3, v_i = 0$, must have $f(v) = 0$. We do so by induction over $k$, where $k$ is the number of voters in $S_1$ with $v_i = 0$. What we proved so far constitutes the base cases of $k = 0, 1$, showing that any $v$ with up to $1$ voter in $S_1$ that has $v_i = 0$, has $f(v) = 0$. We now show that if this holds for some $k$, then it also holds for $k + 1$. Take a vector $v$ that has for some $S' \subseteq S_1$ of size $k+1$, $\forall i \in S_1 \setminus S', v_i = 1$, and $\forall i \in S' \cup S_2 \cup S_3, v_i = 0$. Now choose two different $i_1, i_2 \in S'$, and consider $v'$ so that $v'_{i_1} = 1$, and all other elements are equal to $v$. Consider $v''$ so that $v''_{i_2} = 1$, and all other elements are equal to $v$. Then, $v', v''$ both have $k$ voters in $S_1$ with $v'_i = 0$. Thus, by the induction assumption $f(v') = f(v'') = 0$. Consider $f'$ so that it is identical to $f$, but has $f'(v') = f'(v'') = 1$. 
The utility of all voters in $S_1 \setminus (S' \cup \{i_1, i_2\})$ improves (as they have $v'_i = v''_i = 1$, and $f'(v') = f'(v'') = 1$). The utility of voters $i_1, i_2$ does not change:

\[
\begin{split}
& u_{i_1}(f') \\
& = u_{i_1}(f) - Pr_{v\sim F}[v = v'] + Pr_{v\sim F}[v = v''] \\
& = u_{i_1}(f) - p^{|S_1| - k}(1-p)^{|S_2| + |S_3| + k} + p^{|S_1| - k}(1-p)^{|S_2| + |S_3| + k} \\
& = u_{i_1}(f),
\end{split}
\]
and similarly for $u_{i_2}(f')$. 

The utility of voters $i \in S_2 \cup S_3 \cup S' \setminus \{i_1, i_2\}$ declines (as they have $v'_i = v''_i = 0$, and $f'(v') = f'(v'') = 1$). Overall, by worst-case tie-breaking, we can choose $\mathbf{c}(f, f') = v$, and have $f(\mathbf{c}(f, f')) = f(v) = 1$, in contradiction to self-maintenance. We thus conclude that $f(v) = 0$. 

\qed \end{proof}

\begin{lemma}
\label{lem:unbiased_voters_edge_S1}
    For unbiased i.i.d. voters, for an SCF $f$ and domain $D_G$, if $S_1 = \emptyset$ and $f$ is self-maintaining, then $f$ is either dictatorship, anti-dictatorship, or the constant $0$. 

\end{lemma}
\begin{proof}
    We consider $S_1 = \emptyset$. Then, for all $1\leq i\leq n$, $u_i(f) \geq \frac{1}{2}$. Condition on the number of voters with utility strictly higher than $\frac{1}{2}$:

    \begin{itemize}
        \item \textit{None}. Then, consider some $v$ so that $f(v) = 1$. Let $W$ be the set of voters in $v$ with $v_i = 1$. Consider the $W$-oligarchy to be $f'$. then any voter in $W$ has utility strictly higher than $\frac{1}{2}$ (as the majority of $W$ outcome is positively correlated with $v_i$), and any voter outside $W$ has utility $\frac{1}{2}$ (as for an outside voter $v_i$ is independent of the majority of $W$). Thus, $\mathbf{c}(f,f') = v$, and the voters will choose to move to $f'$. 

        \item \textit{One}. Consider voter $i$ so that $u_i(f) > \frac{1}{2}$. It must be that there is a $v$ with $v_i = 1$ and $f(v) = 1$ (the voter can gain at most $\frac{1}{2}$ in utility from cases where $v_i = 0$ and $f(v) = 0$, with unbiased i.i.d. distribution). All other voters are indifferent between $f$ and an $i$-dictatorship (denote it $f'$), and $i$ strictly prefers it, so $\mathbf{c}(f, f') = v$. 

        \item \textit{Two or more}. 

        Consider any voter $i$ so that $u_i(f) > \frac{1}{2}$. There must be some two vectors $v, v'$ so that $v_i = v'_i$, and $v_{-i} = \neg v'_{-i}$, so that $f(v) = f(v') = v_i$, or otherwise voter $i$'s utility does not surpass $\frac{1}{2}$ (by a one-to-one matching argument). Consider $f'$ so that $f'(v) = \neg v_i, f'(v') = \neg v_i$, and for any other vector $f' = f$. Voter $i$ strictly prefers $f$ over $f'$, and any other voter is indifferent between $f$ and $f'$. Thus, for any $\tilde{v}$ so that $\tilde{v}_i = 0$, it must be that $f(\tilde{v}) = 0$, or otherwise we could use worst-case tie-breaking to set $\mathbf{c}(f,f') = \tilde{v}$ to contradict that $f$ is self-maintaining. Thus, every voter with utility strictly higher than $\frac{1}{2}$ is a veto player. 

        If we denote $H$ as the set of voters with utility strictly higher than $\frac{1}{2}$, the above characterization implies that $f$ performs at best for any $i\in H$ with $u_i(f) \leq Pr[v_i = 0] \cdot 1 + Pr[v_i = 1] \cdot Pr[\forall j \in H \setminus \{i\}, v_j = 1] = \frac{1}{2} + \frac{1}{2^{|H|}}$. Moreover, there must be a winning coalition $S$ so that $H \subseteq S$. Otherwise, $u_i(f) \leq Pr[v_i = 0] \cdot 1 + Pr[v_i = 1] \cdot Pr[\forall j \in H \setminus \{i\}, v_j = 1] \cdot 0 = \frac{1}{2}$, in contradiction to $u_i(f) > \frac{1}{2}$. 

        Let $\eta(v, S', b)$ denote the number of voters $j$ in the set $S'$ that have $v_j = b$. 
        
        If $|H| \geq 3$ is odd, then for the $H$-oligarchy $f'$, for every $i\in H$, 
        \[
        \begin{split}
            & u_i(f') \\
            & = Pr_v[\eta(v, H \setminus \{i\}, v_i) \geq \frac{|H| - 1}{2}] \\
            & = \frac{1}{2^{|H|-1}} \sum_{k=0}^{\frac{|H| - 1}{2}} \binom{|H| - 1}{k} \\
            & \geq \frac{1}{2} + \frac{1}{2^{|H|}} \binom{|H| - 1}{\frac{|H| - 1}{2}} \\
            & \geq u_i(f).
        \end{split}
        \]
        
        If $|H| \geq 3$ is even, then for the $H$-oligarchy $f'$, for every $i\in H$, 
        \[
        \begin{split}
        & u_i(f') \\
        & = Pr[v_i = 0] \cdot Pr_v[\eta(v, H \setminus \{i\}, 0) \geq \frac{|H|}{2} - 1] + Pr[v_i = 1] \cdot Pr_v[\eta(v, H \setminus \{i\}, 1) \geq \frac{|H|}{2} + 1] \\
        & = \frac{1}{2} + \frac{1}{2^{|H|}} \binom{|H| - 1}{\frac{|H|}{2} - 1} \\
        & > u_i(f). 
        \end{split}
        \]
        
        If $|H| = 2$, then for the $H$-oligarchy $f'$, for the two voters $i\in H$, $u_i(f') = \frac{3}{4} \geq u_i(f)$. 

In all the above cases, since for any $i\not \in H$, $u_i(f') = u_i(f) = \frac{1}{2}$, then by worst-case tie-breaking, the voters playing to move to $f'$ can correspond to the winning coalition $S$, and $f$ is not self-maintaining.
     \end{itemize}
\qed \end{proof}

\begin{lemma}
\label{lem:unbiased_voters_edge_S3}
    For unbiased i.i.d. voters, for an SCF $f$ and domain $D_G$, if $S_3 = \emptyset$ and $f$ is self-maintaining, then $f$ is either dictatorship, anti-dictatorship, or the constant $0$. 

\end{lemma}
\begin{proof}
    Consider $S_3 = \emptyset$. Then, for all $1\leq i\leq n$, $u_i(f) < \frac{1}{2}$. Then, for any voter $i$, there must be some two vectors $v, v'$ so that $v_i = v'_i$, and $v_{-i} = \neg v'_{-i}$, so that $f(v) = f(v') = \neg v_i$, or otherwise voter $i$'s utility does not go below $\frac{1}{2}$ (by a one-to-one matching argument). Consider $f'$ so that $f'(v) = f'(v') = v_i$, and for any other vector $f' = f$. Voter $i$ strictly prefers $f'$ over $f$, and any other voter is indifferent between $f$ and $f'$. Thus, for any $\tilde{v}$ so that $\tilde{v}_i = 1$, it must be that $f(\tilde{v}) = 0$, or otherwise we could use worst-case tie-breaking to set $\mathbf{c}(f,f') = \tilde{v}$ to contradict that $f$ is self-maintaining. Since this holds for all the voters, and $f$ is not the identity $0$ function, we conclude that $f(\mathbf{0}) = 1$, and is $0$ otherwise. 

   Now consider the two vectors, $v^1 = (1, 0, \ldots, 0)$, and $v^2 = (0, 1, 0, \ldots, 0)$. We have $f(v^1) = f(v^2) = 0$. Consider $f'$ so that $f'(v^1) = f'(v^2) = 1$, and $f' = f$ otherwise. Then, for all voters, $f'$ is (weakly) worse than $f$, and by worst-case tie-breaking we can choose $\mathbf{c}(f,f') = \mathbf{0}$, and so $f$ is not self-maintaining.

\qed \end{proof}

\UnbiasedVoters*

\begin{proof}

The proof for $D_S$ appears in the main text. 

For the domain $D_G$, we first recall that \textit{unbiased} voters are such that have $p = \frac{1}{2}$. In this case, $2p(1-p) = p^2 + (1-p)^2$, and so $S_2 = \emptyset$. 


Theorem~\ref{thm:dictatorship} shows that $f$ being one of the specified rule types results in it being self-maintaining. We thus consider, going forward, that $f$ is some other SCF. 

In the following, we consider $S_1 \neq \emptyset, S_3 \neq \emptyset$. The case where $S_1 = \emptyset$ or $S_3 = \emptyset$ is covered by Lemma~\ref{lem:unbiased_voters_edge_S1} and Lemma~\ref{lem:unbiased_voters_edge_S3} in the appendix. 

Let $v$ be any vector in $\{0,1\}^n$. We use the characterization of Lemma~\ref{lem:main_structural} to show that $f(v) = 0$. 
First, if $v$ falls within the lemma's characterization, then $f(v) = 0$. Otherwise,
Consider the following two vectors, which are distinct from $v$: Let $v'$ be equal to $v$ for all voters in $S_1$, but have $v'_i = 0$ for all $i\in S_2 \cup S_3$. Let $v''$ be equal to $v$ for all voters in $S_2 \cup S_3$, but have $v''_i = 1$ for all $i\in S_1$. Consider $f'$ so that it is identical to $f$, but has $f'(v') = f'(v'') = 1$. 

The utility of all voters in $S_1$ that have $v_i = 1$ increases, as $v'_i = v''_i = 1$, and we have $c_i(f, f') = 1$. The utility of all voters in $S_2 \cup S_3$ that have $v_i = 0$ decreases, as $v'_i = v''_i = 0$, and we have $c_i(f, f') = 0$. 

For a voter in $S_1$ that has $v_i = 0$, the utility does not change. That is since it has $v'_i = 0, v''_i = 1$, and so:

$$u_i(f') = u_i(f) - Pr_{\hat{v}\sim F}[\hat{v} = v'] + Pr_{\hat{v}\sim F}[\hat{v} = v''] = u_i(f) - \frac{1}{2^n} + \frac{1}{2^n} = u_i(f).$$

For a voter in $S_2 \cup S_3$ that has $v_i = 1$, the utility does not change. That is since it has $v'_i = 0, v''_i = 1$, and so:

$$u_i(f') = u_i(f) - Pr_{\hat{v}\sim F}[\hat{v} = v'] + Pr_{\hat{v}\sim F}[\hat{v} = v''] = u_i(f) - \frac{1}{2^n} + \frac{1}{2^n} = u_i(f).$$

Overall, we conclude that by worst-case tie-breaking we can set $\mathbf{c}(f,f') = v$, and so by the self-maintenance constraint, must have $f(v) = f(\mathbf{c}(f,f')) = 0$. 

\qed \end{proof}

\section{Full Forward-Looking Computational Results}

For $n=3$, due to the small state-space, the forward-looking self-maintaining SCFs are exactly the $10$ self-maintaining with SQP tie-breaking SCFs (the dictatorships, the $3$-oligarchies, and the quasi-dictatorships). The remaining SCF is Unanimity, which is transient.

For $n=4$, a much more interesting outcome emerges. Among the $80$ SCFs in the domain, $30$ are self-maintaining with SQP tie-breaking, but $10$ more are forward-looking self-maintaining. All other SCFs are transient. The SCFs that are self-maintaining with SQP tie-breaking in this case are again the dictatorships, the $3$-oligarchies, and the quasi-dictatorships (the same as Theorem~\ref{thm:opt_SQP}'s pessimistic characterization, without any other SCFs that are self-maintaining only for this distribution). There are two types of SCFs that are forward-looking self-maintaining but not self-maintaining with SQP tie-breaking: (i) The minimal winning coalitions are: $\{ \{1,2,3\}, \{1,2,4\}, \{3,4\}$ (or some permutation of it over the voters), with utilities $(\frac{5}{8}, \frac{5}{8}, \frac{3}{4}, \frac{3}{4})$, (ii) The minimal winning coalitions are: $\{ \{1,2,3\}, \{1,2,4\}, \{1,3,4\}$ (or some permutation of it over the voters), with utilities $(\frac{3}{4}, \frac{5}{8}, \frac{5}{8}, \frac{5}{8})$. Both types beat any of the self-maintaining with SQP tie-breaking SCFs in terms of their utility for the least satisfied voter. Type (i) beats any of the self-maintaining with SQP tie-breaking SCFs in terms of its Nash welfare, whereas $3$-oligarchy has the second best Nash welfare. Type (i) and $3$-oligarchy are equal in terms of their social welfare. 

For $n=5$, there are $80$ self-maintaining with SQP tie-breaking SCFs (as in the $n=4$ and $n=3$ case, these are the dictatorships, the $3$-oligarchies and the quasi-dictatorships). There are $86$ more SCFs that the process identifies as forward-looking self-maintaining, of five types (up to voter permutation). Among these types, the type (up to voter permutations) 
$$\{ \{1,2,5\}, \{1,3,5\}, \{1,4,5\}, \{2,3,5\}, \{2,4,5\}, \{3,4,5\}, \{1,2,3,4\}\}$$ has utilities $(0.625, 0.625, 0.625, 0.625, 0.8125)$, and has strictly higher Nash welfare and strictly higher social welfare than any of the SCFs that are self-maintaining with SQP tie-breaking. 

\begin{figure}[htbp]
    \centering
    \begin{minipage}{0.48\textwidth}\includegraphics[width=\linewidth]{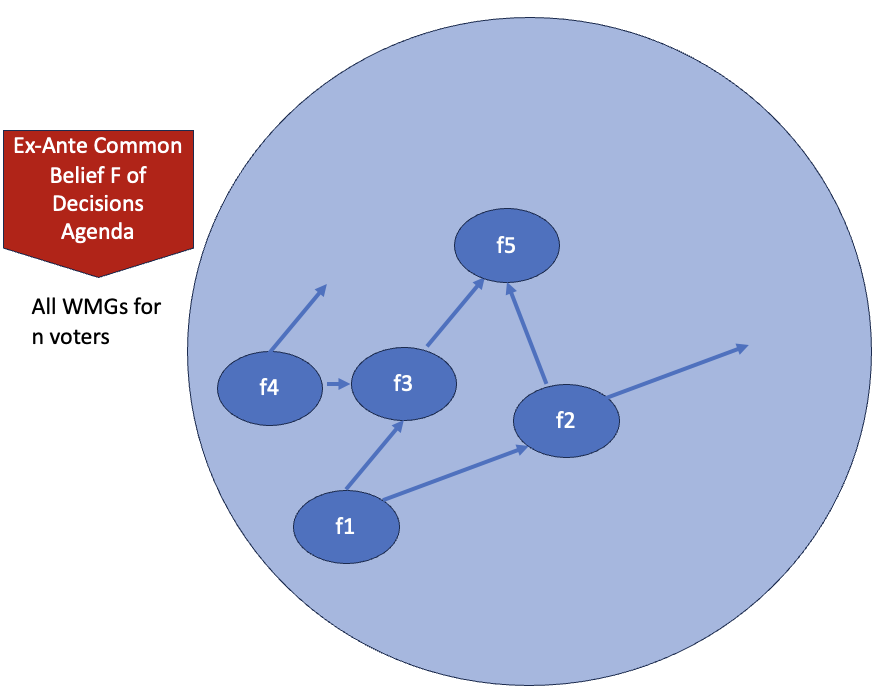}
    \caption{A graphic representation of the stability dynamics. Any SCF in the domain is a node in the graph. A directed edge $f_0 \rightarrow f_1$ exists if $f_0$ loses in the stage $1$ vote of the game of Figure~\ref{fig:self_maintaining_game} when faced with $f_1$. }
\label{fig:forward_looking_dag}
    \end{minipage}\hfill
\begin{minipage}{0.48\textwidth}

\includegraphics[width=\linewidth]{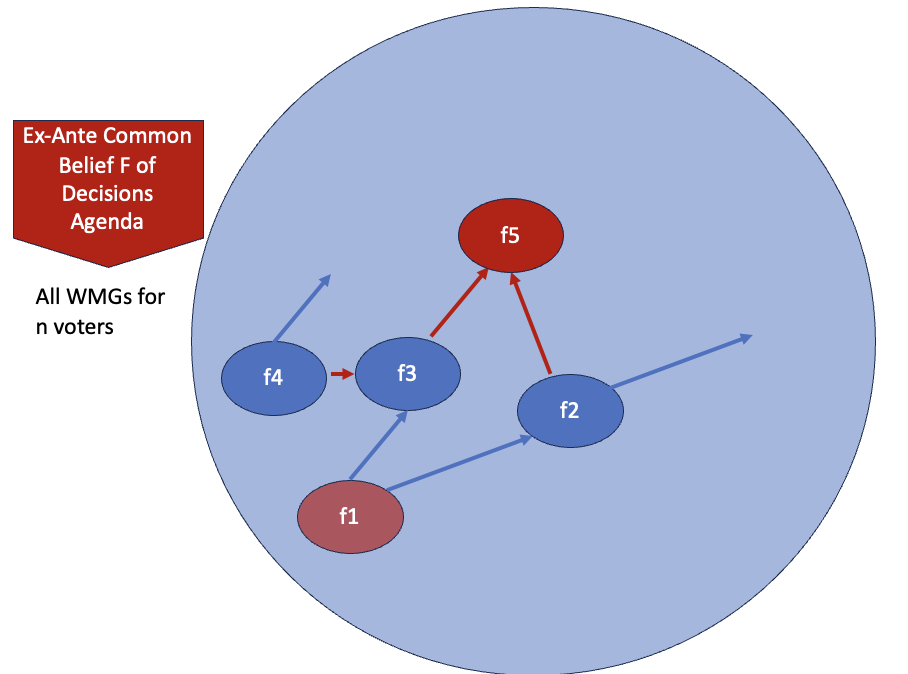}
    \caption{An equilibrium of the forward-looking game. Red nodes are self-maintaining in the equilibrium. A red arrow from (e.g.) $f_4$ to $f_3$ means that $t(f_4) = f_3$. Notice that $f_5$ has no outgoing edges, so it is self-maintaining with SQP tie-breaking. $f_1$ is \textit{not} self-maintaining with SQP tie-breaking, but since in the equilibrium $f_2 \rightarrow f_5, f_3 \rightarrow f_5$, and in the graphic representation $f_1\rightarrow f_2, f_1 \rightarrow f_3$, but not $f_1 \rightarrow f_5$, its only possible choices are $f_2, f_3$, which is not good for $f_1$, because it would yield $t(f_1) = f_5$. We conclude in this equilibrium it must hold that $f_1 \rightarrow f_1$.}
\label{fig:forward_looking_eq}
        \end{minipage}
\end{figure}

\section{Missing Proofs for Bi-Level Voting}

\PessimisticParetoEfficient*

\begin{proof}

(Self-maintaining $\implies$ Pessimistic Pareto-Efficient)

For any distribution $F$ and any alternative SCF $f'$, we have $Unanimity(c(f, f')) = 0$. Thus, there is at least one voter so that $u_i(f, F) \geq u_i(f', F)$, which yields Pessimistic Pareto-Efficient.

(Pessimistic Pareto-Efficient $\implies$ Self-maintaining)

For any distribution $F$ and any alternative SCF $f'$, we have at least one voter so that $u_i(f, F) \geq u_i(f', F)$. This voter has $c_i(f, f') = 0$ by SQP tie-breaking, and so $Unanimity(c(f,f')) = 0$, satisfying pessimistic self-maintenance. 
\qed \end{proof}

\ThreePartition*

\begin{proof}
    Consider the belief $F = UNI([v_{1,2},v_{1,3}, v_{2,3}])$. Consider $f'$ with minimal winning coalitions $v_{1,2}, v_{1,3}, v_{2,3}$. This can be represented as a WMG, for example, by assigning a weight of $|S_1| \cdot |S_2|$ to every voter in $S_3$, $|S_1| \cdot |S_3|$ to every voter in $S_2$, and $|S_2| \cdot |S_3|$ to every voter in $S_1$, and setting $T = 2 \cdot |S_1| \cdot |S_2| \cdot |S_3| -1$.  
    
    Then, for any voter, $u_i(f, F) = \frac{1}{3} < \frac{2}{3} = u_i(f', F)$. We thus have $\mathbf{c}(f, f') = \mathbf{1}$, and since $\hat{f}$ is a simple game, it holds that $\hat{f}(\mathbf{c}(f, f')) = \hat{f}(\mathbf{1}) = 1$. 
\qed \end{proof}

\WMGThresh*

\begin{proof}
    We take advantage of Proposition~\ref{prop:three_partition}. We construct the $3$-partition by a bag-filling process. We first add voters to $S_1$, until their total weight is at least $\frac{W}{3}$. We then add voters to $S_2$ until their total weight is at least $\frac{W}{3}$, and we assign all the remaining voters to $S_3$. By the bag-filling construction, the maximal total weight in voters of any of the sets is less than $\frac{W}{3} + \mu$. Through arithmetics, our condition $\mu \leq \frac{T}{2} - \frac{W}{3}$ is equivalent to $2 (\frac{W}{3} + \mu) \leq T$. Thus, the total weights of any two sets $S_i, S_j$ are not enough to be above the threshold, and $f(v_{i,j}) = 0$. 
\qed \end{proof}

\DecisiveConstitution*

\begin{proof}
    Fix any belief $F$ and alternative SCF $f'$. Since $f \neq f'$, there are some $k$ preference vectors $v_1, \ldots, v_k$ so that $f(v_j) \neq f'(v_j)$. Let the vectors appear with probabilities $p_1, \ldots, p_k$ under $F$. Let $\alpha_{i,j} = \begin{cases} 1 & (v_j)_i = f'(v_j) \\ -1 & (v_j)_i = f(v_j)\end{cases} $ for $1\leq i \leq n, 1\leq j \leq k$. Then, $u_i(f', F) - u_i(f, F) = \sum_{j=1}^k \alpha_{i,j} p_j$. 
    
    If we assume towards contradiction that each voter strictly prefers $f'$ over $f$ (i.e., $u_i(f', F) - u_i(f,F) > 0$), it must be that:

$$\sum_{i=1}^n w_i \cdot \sum_{j=1}^k \alpha_{i,j} p_j > \sum_{i=1}^n w_i \cdot 0 = 0.$$

However, by changing the order of summation, this yields:

$$\sum_{j=1}^k p_j \sum_{i=1}^n w_i \cdot  \alpha_{i,j}  > 0.$$

But this is impossible, since each summand $\sum_{i=1}^n w_i \cdot  \alpha_{i,j}$ is negative: As $f'$ takes preferences with weighted majority according to $w_i$ and changes them to the weighted minority (according to $w_i$). 
    
\qed \end{proof}

\section{The Biased I.I.D. Case}

We go beyond the symmetric assumption that voters are \textit{unbiased}, i.e., $p = \frac{1}{2}$. We show that more self-maintaining SCFs emerge once we relax this assumption. We fully characterize the case of $n=3$. Interestingly, all self-maintaining SCFs with $n=3$ satisfy that their \textit{welfare}, as well as \textit{Nash welfare}, is at most that of a dictatorship. It is plausible that this property holds with general $n$, but for now we leave it as a conjecture. 

\begin{definition}
The \textit{social welfare} of an SCF $f$ with distribution $F$ is $SW(f) \sum_{i=1}^n u_i(f).$

The \textit{Nash welfare} of an SCF $f$ with distribution $F$ is $NW(f) = \pi_{i=1}^n u_i(f)$. 

\end{definition}

Social welfare is a standard way to measure the \textit{efficiency} of an outcome, while the Nash welfare \cite{nsw} has proven to be a good measurement for the \textit{fairness} of the outcome \cite{unreasonableNSW}. 

\begin{restatable}[]{theorem}{iidBiased}
\label{thm:iidBiased}
With $n=3$, and i.i.d. voters with $0 < p < 1$ and $p \neq \frac{1}{2}$, there \textit{are} SCFs outside the characterization of Theorem~\ref{thm:unbiased_voters_char} that are self-maintaining with worst-case tie-breaking. However, there is no non-constant self-maintaining SCF with a higher welfare, or Nash welfare, than a dictatorship. 
\end{restatable}

\begin{proof}

We begin with proving the existence of self-maintaining rules outside the characterization of Theorem~\ref{thm:unbiased_voters_char}. 

The core idea relies on the fact that for any SCF $f$ and voters $i \neq j$, $u_i(f) - u_j(f) \leq \Pr[v_i \neq v_j] = 2p(1-p)$, with equality holding if and only if $f(v) = v_i$ whenever $v_i \neq v_j$.

Consider two cases based on $p$:
\begin{itemize}
    \item \textit{Change-averse ($0 < p < 1/2$):} Let $f(v) = v_3 \land \neg(v_1 \land v_2)$.
    \item \textit{Change-inclined ($1/2 < p < 1$):} Let $f(v) = \neg v_1 \land (v_2 \lor v_3)$.
\end{itemize}

Neither function is constant, a dictatorship, or an anti-dictatorship. We first prove self-maintenance for the change-averse case.

By construction, $f$ always sides with voter $3$ when they disagree with 
voter $1$, attaining the maximal difference: $u_3(f) - u_1(f) = 2p(1-p)$. Assume by contradiction there exists a distinct replacement $f'$ and a ballot $\mathbf{c}$ admissible under worst-case tie-breaking such that $f(\mathbf{c}(f,f')) = 1$. By the definition of $f$, $c_3 = 1$ and at least one of $c_1, c_2$ is $0$. Assume w.l.o.g.\ $c_1 = 0$.

Since $c_3 = 1$ and $c_1 = 0$, we have $u_3(f') \geq u_3(f)$ and $u_1(f') \leq u_1(f)$. Consequently, $2p(1-p) \geq u_3(f') - u_1(f') \geq 2p(1-p) = u_3(f) - u_1(f)$. Together, our inequalities force $u_3(f') = u_3(f)$ and $f'$ must match $f$ on all profiles where $v_1 \neq v_3$.

For the remaining four profiles where $v_1 = v_3$, their point probabilities are $p^3, p^2(1-p), p(1-p)^2$, and $(1-p)^3$. On these profiles, $f$ disagrees with voter $3$ exclusively on the profile $111$, resulting in a utility loss of exactly $p^3$ relative to perfect agreement. Because $p < 1/2$, $p^3$ is strictly smaller than each of the other three probabilities. Therefore, no other sub-collection of these profiles can sum to $p^3$. For $f'$ to achieve the exact same expected utility as $f$, it must disagree with voter $3$ exclusively on $111$. This implies $f' = f$, in contradiction to the assumption that $f'$ is a distinct replacement.

Similarly, for the change-inclined case, by construction, $f$ always sides with voter $2$ when they disagree with voter $1$, attaining the maximal difference: $u_2(f) - u_1(f) = 2p(1-p)$. Assume by contradiction there exists a distinct replacement $f'$ and a ballot $\mathbf{c}$ admissible under worst-case tie-breaking such that $f(\mathbf{c}(f,f')) = 1$. By the definition of $f$, $c_1 = 0$ and at least one of $c_2, c_3$ is $1$. Assume w.l.o.g.\ $c_2 = 1$.

Since $c_2 = 1$ and $c_1 = 0$, we have $u_2(f') \geq u_2(f)$ and $u_1(f') \leq u_1(f)$. Consequently, $2p(1-p) \geq u_2(f') - u_1(f') \geq 2p(1-p) = u_2(f) - u_1(f)$. Together, our inequalities force $u_2(f') = u_2(f)$ and $f'$ must match $f$ on all profiles where $v_1 \neq v_2$.

For the remaining four profiles where $v_1 = v_2$, their point probabilities are $p^3, p^2(1-p), p(1-p)^2$, and $(1-p)^3$. On these profiles, $f$ agrees with voter $2$ exclusively on the profile $000$, resulting in a utility gain of exactly $(1-p)^3$ relative to perfect disagreement. Because $p > 1/2$, $(1-p)^3$ is strictly smaller than each of the other three probabilities. Therefore, no other sub-collection of these profiles can sum to $(1-p)^3$. For $f'$ to achieve the exact same expected utility as $f$, it must agree with voter $2$ exclusively on $000$. This implies $f' = f$, in contradiction to the assumption that $f'$ is a distinct replacement.

Next, we prove the welfare, and Nash welfare, part of the theorem. 

Consider our partition in Definition~\ref{def:IIDpartition} into $3$ sets based on voter utilities. Then, any voter in $S_1, S_2$ has lower utility than a voter $j\neq i$ under $i$-dictatorship. Therefore, if there is more than a single voter in $S_1, S_2$, we have a 1-to-1 matching from each voter to a voter that has (weakly) higher utility under dictatorship (namely the voters \textit{not} in $S_3$ match with a non-dictator, and the remaining voter matches with the dictator). This then implies the theorem both for the sum (social welfare) and product (Nash welfare) of these utilities. 

We thus focus our attention on the case where $|S_3| \geq 2$. 

We first consider $|S_3| = 3$. In this case, it must be that $f(000) = 0$. Otherwise, let $f'$ be the $1$-anti-dictatorship. Ve have $u_1(f') = 0, u_2(f') = u_3(f') = 2p(1-p) < p^2 + (1-p)^2$ for $p \neq \frac{1}{2}$. Since all voters are in $S_3$ under $f$, they have at least utility $p^2 + (1-p)^2$, and so all voters are worse off under $f'$, and we would have $\mathbf{c}(f, f') = 000$, so by self-maintenance it must be that $f(000) = 0$. 

It must also hold that for any $v$ with a single voter $i$ so that $v_i = 1$, we have $f(v) = 0$. Otherwise, we consider $f'$ to be the $i$-dictatorship. We have $c_i(f, f') = 1$, and for the other voters $j \neq i$, $u_j(f') = p^2 + (1-p)^2 \leq u_j(f)$ (since they are in $S_3$). By worse-case tie-breaking, we can choose $c_j(f, f') = 0$, and so having $f(001) = 1$ would contradict self-maintenance. 

This further implies, for any $v$ with exactly two voters $i,j$ so that $v_i = v_j = 1$, it must hold that $f(v) = 0$. Otherwise, let $v^i$ be the voting preference where voter $i$ is the unique voter with $v_i = 1$, and $v^j$ similarly the voting preference where voter $j$ is the unique voter with $v_j = 1$. Let $f'$ be the SCF that is the same as $f$ other than $f'(v^i) = f'(v^j) = 1$. It holds that $u_i(f) = u_i(f'), u_j(f) = u_j(f')$, and for the remaining voter $i'$, it holds that $u_{i'}(f) > u_{i'}(f')$. By worst-case tie-breaking, we can choose $c_i(f, f') = c_j(f, f') = 1, c_{i'}(f, f') = 0$, and so by self-maintenance it must hold that $f(v) = 0$.   

Finally, the only remaining non-constant $f$ is Unanimity, which has $u_i(f) = 1 - p + p^3$ for all voters $i$. All voters prefer Majority (denote it $f'$) which has $u_i(f') = 1 - p(1-p) = 1 - p + p^2 > 1 - p + p^3$. 

We next consider the case $|S_3| = 2$. W.l.o.g., we let voter $1$ be in $S_1 \cup S_2$, and voters $2$ and $3$ are in $S_3$. 

We note the following. 

If $f(100) = 1$, then the SCF is not self-maintaining, as it would move to a dictatorship by voter $1$: voter $1$ improves from being in $S_1$ to having utility $1$, and voters $2$ and $3$ do not improve because their utility from $1$-dictatorship is $p^2 + (1-p)^2$, weakly less than their utility, as they are in $S_3$. 

If $f(110) = 1$, then the SCF is not self-maintaining, as it would move to a dictatorship by voter $2$: Voter $1$ would improve from being in $S_1$ to utility $p^2 + (1-p)^2$, voter $2$ would improve to $1$ from a utility lower than $1$ (which it has under any SCF other than $2$-dictatorship with $0 < p < 1$, as there is non-zero probability to draw a vector where its preference does not match the outcome), and voter $3$ would not improve as $p^2 + (1-p)^2$ under $2$-dictatorship is weakly less than their utility, as they are in $S_3$. 

Similarly to the last case, if $f(101) = 1$, then the SCF is not self-maintaining, as it would move to a dictatorship by voter $3$. If $f(000) = 1$, it would move to an anti-dictatorship of agent $1$: Voters $2$ and $3$ would have utility $2p(1-p) \leq p^2 + (1-p)^2$, weakly worse than their utility as they are in $S_3$, and agent $1$ would have utility $0$. 

If $f(111) = 1$, notice it must be (by the previous cases) that $f(101) = f(110) = 0$. Consider $f'$ which is identical to $f$ on all preferences besides $f'(101) = f'(110) = 1$. For voter $1$, $f'$ is an improvement, and the other two voters are indifferent between $f$ and $f'$, and so by worst-case tie-breaking may vote to move away. 

All in all, we conclude that $f(000) = f(100) = f(101) = f(110) = f(111) = 0$. 
This leaves three vectors with unspecified outcome under $F$, and thus $2^3 = 8$ possible values for $f$. 
We thus continue with a case analysis for the three remaining vectors. 

\begin{itemize}
\item If $f(001) = f(010) = 0$. Then, it must be that $f(011) = 0$. This is since we may consider $f'(001) = f'(010) = 1$, and for any other $v$, $f'(v) = f(v)$. We have $Pr_{v\sim F}[v = 001] = Pr_{v\sim F}[v = 010] = p(1-p)^2$, and by worst-case tie-breaking, 
$\mathbf{c}(f, f') = 011$. Then, by the self-maintenance constraint, it must be that $f(\mathbf{c}(f, f')) = f(011) = 0$. 
We conclude that in this case $f = 0$ is constant. 

\item If $f(011) = 0$, since we know that $f(101) = 0$ (from the preceding analysis), consider $f'$ which is identical to $f$ other than $f'(011) = f'(101) = 1$. It has the same utility for voters $1$ and $2$, and higher utility for voter $3$, and so it must be that $f(001) = 0$, or we would have $f(\mathbf{c}(f, f')) = f(001) = 1$, in contradiction to self-maintenance. 

Also, since we know that $f(110) = 0$ (from the preceding analysis), we can similarly consider an $f'$ identical to $f$ but with $f'(011) = f'(110) = 1$. It has the same utility for voters $1$ and $3$, and higher utility for voter $2$, and so it must be that $f(010) = 0$. 

Since $f(010) = f(001) = 0$, we are back in the previous case analysis. 

\item $f(011) = 1, f(001) = 1, f(010) = 0$. The voter utilities in this case are $(1-p)^2, (1-p)(p + p^2 + (1-p)^2), (1-p)(1+p)$, respectively. It can be directly shown that there is no $p$ for which the sum of these utilities is higher than the dictatorship welfare, and no $p$ for which the product of these utilities is higher than the dictatorship Nash welfare. We show that $f$ is self-maintaining: Notice that to violate self-maintenance, it must be that $f'$ is (weakly) preferred by voter $3$, and (weakly) not preferred by voter $1$. However, the vectors that do not follow voter $3$ preference are $f(111) = 0, f(101) = 0$, where voter $3$ agrees with voter $1$. Similarly, the vectors that \textit{do} follow voter $1$ preferences are $f(000) = 0, f(010) = 0$. All in all, it holds that for any $f'$, $u_3(f') - u_3(f) \leq u_1(f') - u_1(f)$. Thus, in order to have $c_3 = 1$, we must also have $c_1 = 1$, but then $f(\mathbf{c}) = 0$. 

\item $f(011) = 1, f(001) = 1, f(010) = 1$. In this case, $f$ is equal to the voter $1$-anti-dictatorship, besides the fact that $f(000) = 0$ while $f^1_{anti-dict}(000) = 1$. 
Thus, $\forall 1\leq i\leq 3, u_i(f) = u_i(f^1_{anti-dict}) + (1-p)^3$. It can be shown directly that both the social welfare and Nash welfare of these utilities under-perform these of the dictatorship for any $0 < p < 1$.  
Regarding self-maintenance, we note that any $v$ with $f(v) = 1$ has $v_1 = 0$. Thus, voter $1$ needs to weakly prefer $f$ over $f'$. However, $f(000) = 0$ is the only vector so that voter $1$ is satisfied with $f$. We note that in this case, it holds that for any $f'$, $u_3(f') - u_3(f) \leq u_1(f') - u_1(f)$ and also $u_2(f') - u_2(f) \leq u_1(f') - u_1(f)$. Thus, if either $c_3 = 1$ or $c_2 = 1$, it must be that also $c_1 = 1$. This implies $f(\mathbf{c}) = 0$. 
\end{itemize}
\qed 
\end{proof}

We now consider our second tie-breaking rule, that of \textit{status-quo precedence}. We can show a construction of an SCF for general $n$, in particular $n=3$ as well, that is self-maintaining and has higher social and Nash welfare than a dictatorship.

\begin{restatable}[]{theorem}{consDuo}
\label{thm:consDuo}
The $2$-oligarchy is self-maintaining with SQP tie-breaking for any $0 < p < 1$. Moreover, it has better social and Nash welfare than a dictatorship with \textit{change-averse} ($p < \frac{1}{2}$) voters. 
\end{restatable}

\begin{proof}

Since $2$-oligarchy is self-maintaining in the pessimistic case with SQP tie-breaking, it particularly is self-maintaining w.r.t. any biased i.i.d. distribution. 

We now consider social and Nash welfare. For any voter $k$ other than $i,j$, their utility under $f^{i,j}_{2-olig}$ has 
\[
\begin{split}
    & u_k(f^{i,j}_{2-olig}) \\
    & = (1-p) \cdot 2p(1-p) + p^3 + (1-p)^3 \\
    & \stackrel{\text{change-averse}}{>} \frac{1}{2} \cdot 2p(1-p) + p^3 + (1-p)^3 \\
    & = p(1-p) + p^3 + (1-p)^3 \\
    & \geq p^2 + (1-p)^2 = u_k(f^i_{dict}). 
    \end{split}
    \]

It is thus a sufficient condition for the Nash welfare and social welfare of $f^{i,j}_{2-olig}$ to surpass $f^i_{dict}$ to have that the product and the sum of $u_i, u_j$ is better under the former. 
We can directly write these conditions:

\[
\begin{split}
    & \left(p^2 + (1-p) \right)^2 \geq p^2 + (1-p)^2,\\
    & 2\left(p^2 + (1-p)\right) \geq p^2 + (1-p)^2 + 1,
\end{split}
\]  

and verify they hold for any $p$. 
\qed \end{proof}

We remark that as a function of $n$, there are some \textit{change-inclined} values of $p > \frac{1}{2}$ where the social and Nash welfare of the $2$-oligarchy surpass that of a dictatorship, as well.

\section{Restricting to Anonymous SCFs}
\label{app:anonymous}

In many natural settings, such as Parliaments, only \textit{anonymous} SCFs are considered. I.e., we may require that a decision reaches consensus, or a qualified majority, or a simple majority, but we do not give preferential treatment to votes by a certain party or another \footnote{This is only true up to a ``first approximation''. There are many cases with special roles for different voters, such as the veto countries in the UN security council, or the US vice-president having a tie-breaking vote in the senate.}. We give an overview of how our results translate to this case. For convenience, we only consider odd $n$ (so that we never have ties). Overall, the conclusion is that simple majority plays a similar role to dictatorship, once anonymity is introduced, in the sense that voters would move away from ``checks and balances'' that they may view as prohibitive.

\begin{definition}
\textit{Anonymous SCF}. We say $f$ is anonymous if for any permutation $\pi$ over the voters, and any $v\in \Omega$, $f(\pi(v)) = f(v)$. 

\textit{Monotone SCF}. We say $f$ is monotone if for any two $v^1, v^2 \in \Omega$ so that $S_1 \subseteq S_2$, if $f(v^1) = 1$ then $f(v^2) = 1$. I.e., a decision may not be rejected because \textit{more} voters vote for it. 
\end{definition}

There are exactly $n$ anonymous, monotone and non-constant SCFs: Anonymous implies that we can encode the SCF as a function $\tilde{f}:[n] \rightarrow \{0,1\}$, where the input is the number of voters who vote $1$. Then, adding monotone (and non-constant) on top of it implies that the SCF takes the form of a threshold function: Some $1\leq k\leq n$ so that for a vector $v$ with a set $S$ of voters with $v_i = 1$, we have $f(v) = 1[|S| \geq k]$. 

It is straight-forward to redefine our different approaches in this setting. Instead of considering the optimistic and pessimistic approach w.r.t. to \textit{all} SCFs, we restrict the space only to anonymous and monotone SCFs.

\begin{theorem}
\label{thm:anonymous_opt}
Limited to anonymous, monotone, and non-constant SCFs with $n$ voters , in the optimistic approach with worst-case tie-breaking, any SCF $f$ is self-maintaining if and only if its threshold $k > \frac{n+1}{3}$. 
\end{theorem}

\begin{proof}

(Self-maintaining implies characterization)

Assume by contradiction that $3k \leq n+1$ and $f_k$ is self-maintaining under some belief $F$. We compare $f_k$ with the adjacent threshold $f_{k+1}$. For any voter $i$, $u_i(f_k) - u_i(f_{k+1}) = \sum_{|S(v)|=k} (\mathbf{1}_{v_i=1} - \mathbf{1}_{v_i=0})F(v)$.

For any set of voters $S$ with $|S| \geq 2k$, exactly $k$ voters vote $1$ in any profile $v$ where $|v|=k$, meaning at least $k$ voters in $S$ vote $0$. Thus, the sum of utility differences over $S$ is $\leq 0$, making it impossible for $2k$ or more voters to strictly prefer $f_k$.

Thus, the set $S$ of voters who strictly prefer $f_k$ has size $|S| < 2k$. Under worst-case tie-breaking, the adversary can assign $c_i = 1$ to all remaining voters (those indifferent or preferring $f_{k+1}$). The number of such $1$ votes on the ballot $\mathbf{c}$ is $n - |S| \geq n - 2k + 1$. Since $3k \leq n+1$, we have $n - 2k + 1 \geq k$. Therefore, $\mathbf{c}$ has at least $k$ affirmative votes, so $f_k(\mathbf{c}) = 1$, in contradiction to self-maintenance.

(Characterization implies self-maintaining)

If $n < 2k$, the quota rule $f_k$ is proper, and by Theorem~\ref{thm:opt_arbitrary}, it is optimistically self-maintaining. We thus focus on the case where $2k \leq n < 3k - 1$.

Let $T$ be a subset of voters with $|T| = n - k + 1$. Let $v^*$ be the profile where $S(v^*) = [n] \setminus T$, meaning exactly the $k-1$ voters outside $T$ vote $1$. Let $U$ be the uniform distribution over the $\binom{n-k+1}{k}$ profiles $v$ with $S(v) \subseteq T$ and $|S(v)| = k$. Consider the belief $F$ which is $v^*$ w.p. $\frac{1}{2}$ and a profile $v$ drawn from $U$ w.p. $\frac{1}{2}$.

For any voter $i \in T$, we have $v_i^* = 0$. Since $|S(v^*)| = k-1 < k$, $f_k(v^*) = 0$, matching $i$'s preference. Under $U$, $f_k$ accepts every profile, matching $i$'s preference when $v_i=1$, which occurs with probability $\frac{\binom{n-k}{k-1}}{\binom{n-k+1}{k}} = \frac{k}{n-k+1}$. Thus, $u_i(f_k, F) = \frac{1}{2} + \frac{1}{2}\left(\frac{k}{n-k+1}\right)$.

Against any alternative threshold rule $f_{\ell}$ with $\ell \neq k$:
\begin{itemize}
    \item If $\ell < k$: $f_{\ell}$ accepts $v^*$ since $k-1 \geq \ell$, disagreeing with $i$. The maximum possible utility for $i$ is bounded by $u_i(f_{\ell}, F) \leq \frac{1}{2} < u_i(f_k, F)$.
    
    \item If $\ell > k$: $f_{\ell}$ rejects $v^*$ (matching $i$). However, $f_{\ell}$ also rejects all profiles in $U$, matching $i$'s preference only when $v_i = 0$, which occurs with probability $\frac{\binom{n-k}{k}}{\binom{n-k+1}{k}} = \frac{n-2k+1}{n-k+1}$. Thus, $u_i(f_{\ell}, F) \leq \frac{1}{2} + \frac{1}{2}\left(\frac{n-2k+1}{n-k+1}\right)$. Because $n+1 < 3k \implies n - 2k + 1 < k$, we have $u_i(f_{\ell}, F) < u_i(f_k, F)$.
\end{itemize}

Consequently, for any $f_{\ell} \neq f_k$, every voter $i \in T$ strictly prefers $f_k$. In any worst-case choice ballot $\mathbf{c}$ generated against $f_{\ell}$, all $n - k + 1$ voters in $T$ must vote $0$. The number of affirmative votes is at most $n - (n - k + 1) = k - 1$. Since $k-1 < k$, $f_k(\mathbf{c}) = 0$, satisfying self-maintenance. 
    
\end{proof}


\begin{theorem}
Limited to anonymous, monotone and non-constant SCFs, in the pessimistic approach with worst-case tie-breaking, no SCF is self-maintaining. 
\end{theorem}

\begin{proof}[Sketch]
 This repeats the same construction as in the general non-anonymous case: We fix $F$ to be some $v$ w.p. $1$, and consider $f'$ that has the same outcome as $f$ over $v$, but not elsewhere. 
\qed \end{proof}

\begin{theorem}
\label{thm:anonymous_pess}
Limited to anonymous, monotone, and non-constant SCFs, in the pessimistic approach with SQP tie-breaking, an SCF $f$ is self-maintaining if and only if it is simple-majority in the case $n=3$.
\end{theorem}
\begin{proof}
As we later see in the i.i.d. case (Theorem~\ref{thm:anonymous_iid}), simple-majority rule is preferred over all other rules for the unbiased i.i.d. belief with SQP tie-breaking. Thus, if we take unbiased i.i.d. to be the pessimistic belief, it rules out all rules other than simple-majority. It is left to show that simple-majority is self-maintaining if and only if $n=3$. 

If $n=3$, then simple-majority is the same as $3$-oligarchy, and Theorem~\ref{thm:pess_SQP_char} shows $3$-oligarchy is self-maintaining in the pessimistic approach with SQP tie-breaking in the general case. In particular, this holds for all anonymous and monotone SCFs.

\begin{table}[htbp]
    \centering
    \renewcommand{\arraystretch}{1.3} 
    \begin{tabular}{lccccccccc}
        \toprule
        & $1$ & $\ldots$ & $k-3$ & $k-2$ & $k-1$ & $k$ & $k+1$ & $\ldots$ & $2k-1$  \\
        \midrule
        $v^1$ & $0$ & $\ldots$ & $0$ & $0$ & $0$ & $1$ & $1$ & $\ldots$ & $1$\\        
        $v^2$ & $0$ & $\ldots$ & $0$ & $0$ & $1$ & $0$ & $1$ & $\ldots$ & $1$\\
        $v^3$ & $0$ & $\ldots$ & $0$ & $1$ & $0$ & $0$ & $1$ & $\ldots$ & $1$\\
        \bottomrule
    \end{tabular}
    \caption{Voting preferences for Theorem~\ref{thm:anonymous_pess}}
    \label{tab:anonymous_pess}
\end{table}

Otherwise, consider $k$ so that $n = 2k - 1$. $k$ corresponds to the minimal amount of votes that constitute a majority. Thus, for any $v$ with $k$ voters who vote $1$, $f_{sm}(v) = 1$. Consider the preference vectors given in Table~\ref{tab:anonymous_pess}. It holds that $f_{sm}(v^1) = f_{sm}(v^2) = f_{sm}(v^3) = 1$. It also holds that for any $k-2\leq i \leq k$, they have $v_i = 0$ for exactly two of the three vectors. Thus, if we choose the belief $F$ so that $\forall 1\leq i\leq 3,Pr_{\hat{v} \sim F}[\hat{v} = v^i] = \frac{1}{3}$, and $f'$ be the qualified majority that requires at least $k+1$ votes for a motion to pass, then $f'(v^1)= f'(v^2) = f'(v^3) = 0$, and $\forall k-2\leq i\leq k$, $u_i(f_{sm}) = \frac{1}{3} < \frac{2}{3} = u_i(f')$. For $1\leq i \leq k -3$, it holds that $u_i(f_{sm}) = 0 < 1 = u_i(f')$. Thus, for all $1\leq i \leq k$, $c_i(f_{sm},f') = 1$, and there is a majority that supports moving from $f_{sm}$ to $f'$, so $f_{sm}(\mathbf{c}(f_{sm}, f')) = 1$, contradicting self-maintenance. 
\qed \end{proof}

\begin{theorem}
\label{thm:anonymous_iid}
Limited to anonymous and monotone SCFs, with i.i.d. beliefs and SQP tie-breaking, for any $0 < p < 1$ (unbiased, change-averse and change-inclined) an SCF is self-maintaining if and only if it is simple-majority.

\end{theorem}

\begin{proof}
\textit{Proof-sketch.} With an anonymous SCF $f$ and i.i.d. belief $F$, the overall symmetry implies that $\forall 1 \leq i \leq n, u_i(f) = \frac{1}{n} SW(f,F)$. Thus, an SCF is self-maintaining if and only if it maximizes welfare for $F$. Since the majority vote maximizes welfare for any $F$ (since for any $v \in \Omega$, it benefits the most voters, as it goes with the majority), and with i.i.d. $F$, every $v\in \Omega$ has a positive probability to occur, and so any other SCF has strictly lower welfare, we conclude that simple-majority is the unique 
self-maintaining SCF. 

\qed \end{proof}

\end{document}